\documentclass[1p,singlecolumn,review]{elsarticle}
\usepackage{graphicx}  %
\usepackage{subcaption}
\usepackage{multirow}
\usepackage{makecell}
\usepackage{booktabs}
\usepackage{amsmath}
\usepackage{siunitx}
\DeclareSIUnit{\charge}{\milli\volt\nano\second}
\DeclareSIUnit{\MeV}{\mega\electronvolt}
\DeclareSIUnit{\days}{days}
\usepackage{bm}        %
\usepackage{enumerate}
\usepackage{stfloats}
\usepackage{color}
\usepackage{appendix}
\usepackage{tabularx}
\usepackage[colorlinks,linkcolor=blue,anchorcolor=blue,citecolor=blue]{hyperref}
\usepackage{tikz}
\usetikzlibrary{shapes.geometric, arrows}
\usepackage{rotating}
\usepackage[version=3]{mhchem}
\usepackage{footnote}
\usepackage{tikz}
\usetikzlibrary{backgrounds,mindmap,calc,positioning,intersections}

\newcommand{\linelabel}[1]{\label{#1}}

\tikzstyle{startstop} = [rectangle, rounded corners, minimum width = 1cm, minimum height=0.5cm,text centered, draw = black]
\tikzstyle{io} = [trapezium, trapezium left angle=70, trapezium right angle=110, minimum width=1cm, minimum height=0.5cm, text centered, draw=black]
\tikzstyle{process} = [rectangle, minimum width=3cm, minimum height=1cm, text centered, draw=black]
\tikzstyle{decision} = [diamond, aspect = 3, text centered, draw=black]
\tikzstyle{arrow} = [->,>=stealth]

\begin{document}

\title{Performance of the 1-ton Prototype Neutrino Detector at CJPL-I\tnoteref{license}}%
\tnotetext[license]{\includegraphics[height=\fontcharht\font`\B]{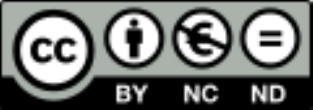} \copyright 2023. This manuscript version is made available under the CC-BY-NC-ND 4.0 license \url{https://creativecommons.org/licenses/by-nc-nd/4.0/}\\ This is an accepted manuscript version of an article, available at \url{https://doi.org/10.1016/j.nima.2023.168400}}

\author[TUDEP,TUHEP]{Yiyang Wu}
\author[TUDEP,TUHEP]{Jinjing Li\corref{cor1}}
\ead{jinjing-li@mail.tsinghua.edu.cn}
\cortext[cor1]{Corresponding author.}
\author[TUDEP,TUHEP,TUPRI]{Shaomin Chen}
\author[TUDEP,TUHEP]{Wei Dou}
\author[TUDEP,TUHEP]{Lei Guo}
\author[TUDEP,TUHEP]{Ziyi Guo}
\author[TUDEP,TUHEP]{Ghulam Hussain}
\author[TUDEP,TUHEP]{Ye Liang}
\author[UCASP]{Qian Liu}
\author[SYSUP]{Guang Luo}
\author[UCASP]{Wentai Luo}
\author[NUSP]{Ming Qi}
\author[TUDEP,TUHEP]{Wenhui Shao}
\author[SYSUP]{Jian Tang}
\author[TUDEP,TUHEP]{Linyan Wan}
\author[TUDEP,TUHEP,TUPRI]{Zhe Wang}
\author[TUDEP,TUHEP,TUPRI]{Benda Xu}
\author[TUDEP,TUHEP]{Tong Xu}
\author[TUDEP,TUHEP]{Weiran Xu}
\author[TUDEP,TUHEP]{Yuzi Yang}
\author[TUDEP,TUHEP]{Lin Zhao}
\author[TUDEP,TUHEP]{Aiqiang Zhang}
\author[TUDEP,TUHEP]{Bin Zhang}

\affiliation[TUDEP]{Department of Engineering Physics, Tsinghua University, Beijing 100084, China}
\affiliation[TUHEP]{Center for High Energy Physics, Tsinghua University, Beijing 100084, China}
\affiliation[TUPRI]{Key Laboratory of Particle & Radiation Imaging (Tsinghua University), Ministry of Education, China}
\affiliation[UCASP]{School of Physical Sciences, University of Chinese Academy of Sciences, Beijing 100049, China}
\affiliation[SYSUP]{School of Physics, Sun Yat-Sen University, Guangzhou 510275, China}
\affiliation[NUSP]{School of Physics, Nanjing University, Nanjing 210093, China}

\date{\today}%

\begin{abstract}
        China Jinping Underground Laboratory provides an ideal site for solar, geo-, and supernova neutrino studies. With a prototype neutrino detector running since 2017, containing 1-ton liquid scintillator, we tested its experimental hardware, performed the detector calibration and simulation, and measured its radioactive backgrounds, as an early stage of the Jinping Neutrino Experiment (JNE). We investigated the radon background and implemented the nitrogen sealing technology to control it. This paper presents the details of these studies and will serve as a key reference for the construction and optimization of the future large detector of JNE.
\end{abstract}


\maketitle

\section{Introduction}
\label{sec:introduction}

The China Jinping Underground Laboratory (CJPL) sits inside Jinping mountain in Sichuan Province, China. With a vertical rock overburden of about \SI{2400}{m}, it becomes one of the world's deepest underground laboratories~\cite{cheng2017china}. The closest nuclear power plant is approximately 1000 km away, giving it the lowest reactor neutrino flux among all the underground laboratories in the world~\cite{Jinping:2016iiq}. These remarkable features make it an ideal site for a low-energy neutrino laboratory.

\begin{figure*}[!htb]
    \centering
    \includegraphics[width=0.85\textwidth]{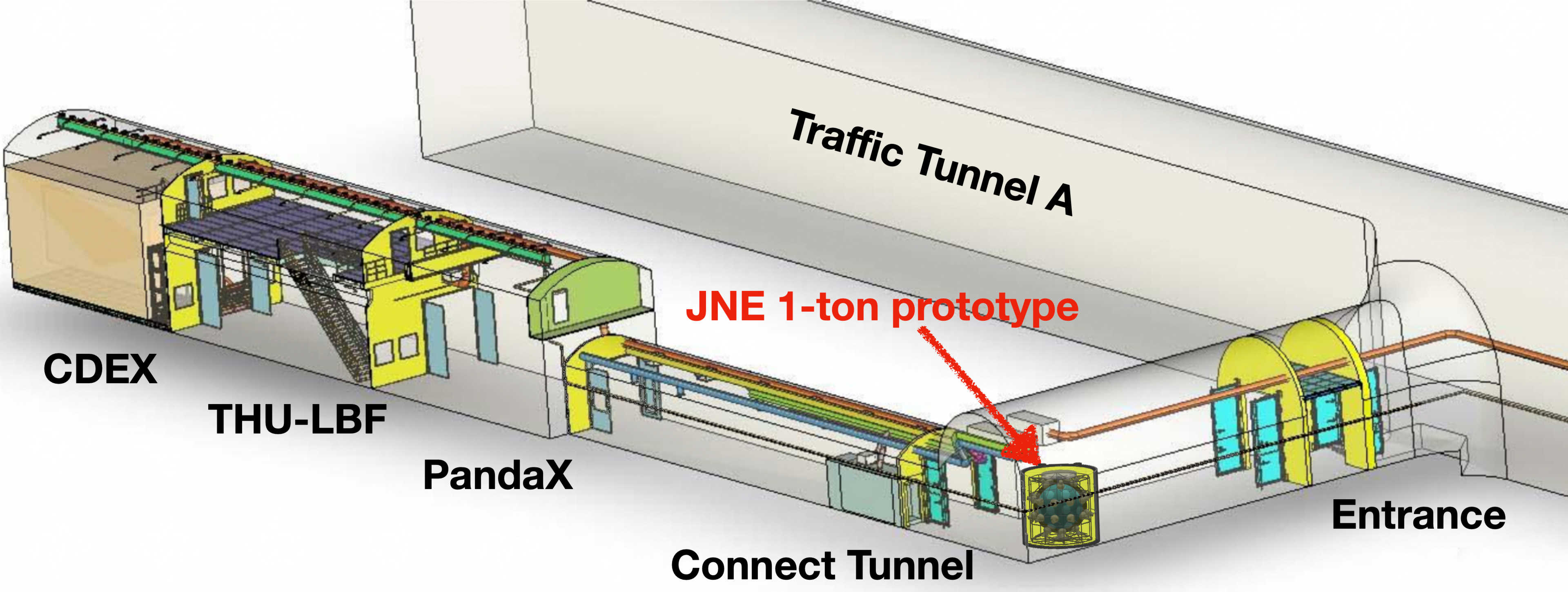}
    \caption{Schematic layout of hall A at CJPL-I, showing the location of the CDEX, PandaX experiments, the low-background screening facilities (THU-LBF), as well as the 1-ton prototype of JNE. The dimensions of the main experiment hall are 6.5 m (width) $\times$ 6.5 m (height) $\times$ 40 m (length). This figure is adapted from Ref.~\cite{Cheng:2017usi}.}
    \label{fig:F1-CJPL1}
\end{figure*}

The first phase of the Jinping laboratory (CJPL-I) was constructed in the middle of the traffic tunnels at the end of 2009. Two dark matter experiments, CDEX~\cite{CDEX:2022rxz} and PandaX~\cite{PandaX-II:2021kai}, are now running at CJPL-I, as shown in Fig.~\ref{fig:F1-CJPL1}. The second phase of the Jinping laboratory (CJPL-II) started at the end of 2014 and will provide space for more underground experiments~\cite{Li:2014rca}.

The proposed Jinping Neutrino Experiment (JNE)~\cite{Jinping:2016iiq} aims to study MeV-scale neutrinos, including solar neutrinos, geo-neutrinos, supernova neutrinos, etc. The experiments targeting MeV-scale neutrinos are prone to contamination from radioactive backgrounds within the detector material. Therefore, a practical measurement of the radioactive backgrounds and a comprehensive implementation of detection techniques, i.e., hardware and software, are necessary to design and optimize future detectors. \linelabel{line:full-scale}A multi-hundred-ton detector is now under construction and will be finished around 2026 to verify further the technologies, to reach the goal of a light yield of 500 photoelectrons\linelabel{line:500pe-goal} (PE) per MeV or better, and\linelabel{line:full-scale-requirements} a contamination level close to that of Borexino~\cite{BOREXINO:2022abl}.

\linelabel{line:1t goal}A 1-ton detector, built as a prototype of the JNE at CJPL-I hall A, has been running since 2017~\cite{Wang:2017ynm}. The location of the prototype is shown in Fig.~\ref{fig:F1-CJPL1}.
The primary goal of the prototype is to systematically test and understand several key detector components, including slow liquid scintillator (LS), photomultiplier tubes (PMTs) and electronics, and to measure the backgrounds. With this prototype, previous studies estimated the cosmic-ray muon~\cite{JNE:2020bwn} and muon-induced background~\cite{JNE:2021cyb} at CJPL-I, enabling a better understanding of the cosmic-ray background in the experiment. The whole process of detector calibration, simulation, and event reconstruction is summarized in this paper. We subsequently analyzed $\beta$-$\alpha$ cascade decays and $\gamma$ signals from the decay chains of natural isotopes, and observed different radon leakage features in different operating stages. These measurements can validate the already used detection techniques and provide references for future detector optimizations.

After detailing the design and calibration of the 1-ton prototype in Sec.~\ref{sec:detector-design}, we present the simulation framework and its validation in Sec.~\ref{sec:detector-simulation}. Then, we show the analysis of the radioactive backgrounds in Sec.~\ref{sec:measurement}. Finally, we give a summary of this research in Sec.~\ref{sec:summary}.

\section{The 1-ton prototype}
\label{sec:detector-design}

\subsection{Detector design and installation}
\begin{figure}[!htb]
    \centering
    \includegraphics[width=0.8\linewidth]{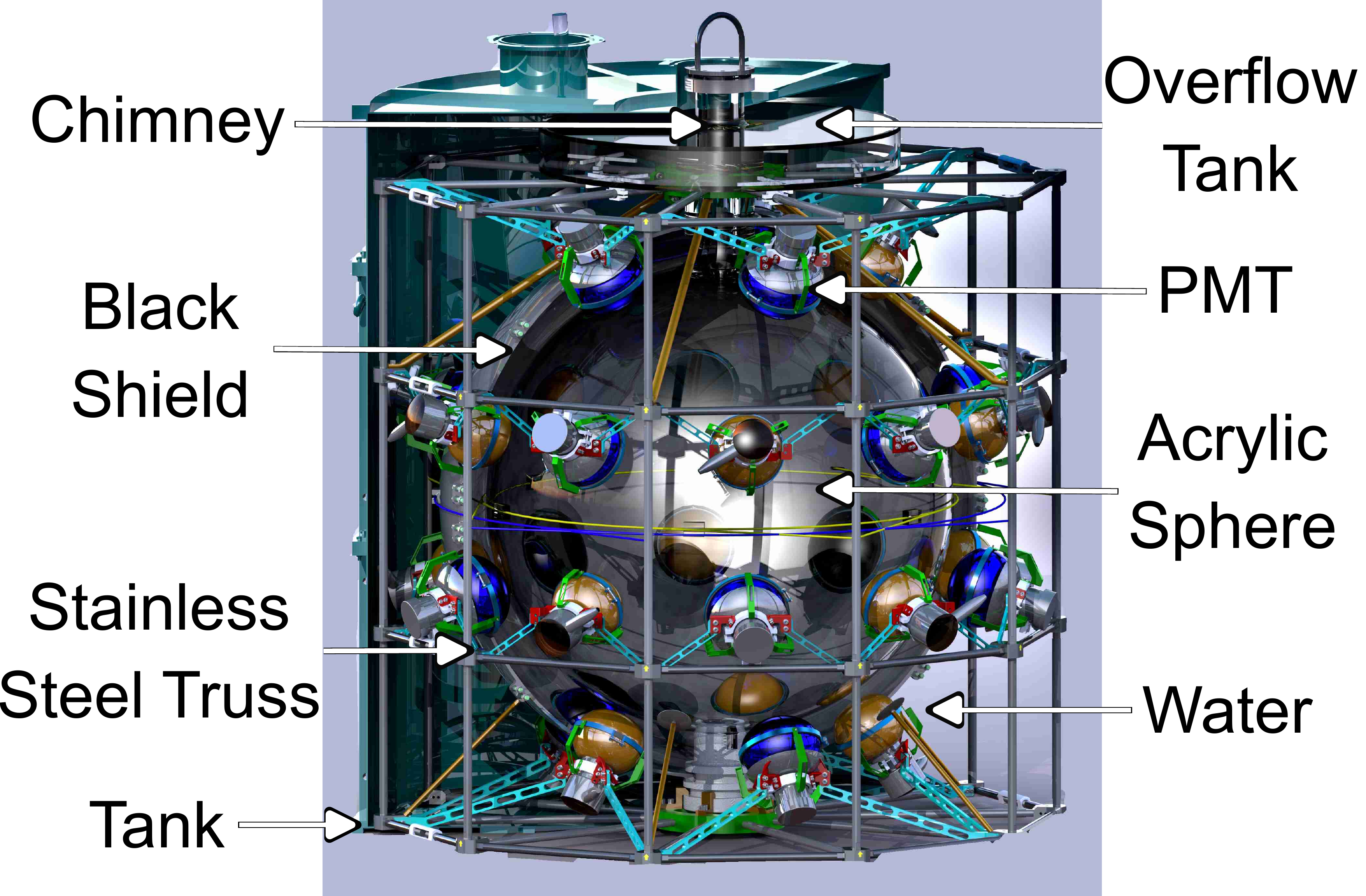}%
    \caption{\label{fig:1tonschema} Schematic figure of the 1-ton prototype for the JNE.}
\end{figure}

\linelabel{line:begin-1ton-description}Fig.~\ref{fig:1tonschema} shows a schematic figure of the 1-ton prototype. The target material, i.e., slow LS candidate~\cite{Li:2015phc,Guo:2017nnr}, is contained by the innermost spherical acrylic vessel with an inner radius of 645 mm and a thickness of 20 mm.
The acrylic vessel comprises three parts: a sphere, a chimney, and an overflow tank. They are jointed with each other through bulk polymerization. The target volume is built to be spherical so that the detector can have a more uniform response than a cylindrical or rectangular one. The chimney is used to fill the liquids into the target volume. The overflow tank is for avoiding overflow at temperature fluctuation. Besides the opening on the top of the acrylic vessel, a pipe is connected to the bottom with one filling outlet on top of the prototype to enable liquid circulation and nitrogen bubbling.

A stainless steel tank holds the acrylic vessel. This 4-mm-thick tank measures 2000 mm in diameter, and 2090 mm in height, filled with pure water during the operation. It has five flanges on the top, which are 114 mm in height and 242 mm in diameter. The middle one is set for the acrylic chimney. Two edge ones are for routing the PMT cables, while the other two are for water filling in the tank, as well as liquid circulation and nitrogen bubbling in the acrylic vessel.

The tank is enclosed by a 5-\unit{\centi\meter}-thick lead wall, severing as a passive shielding together with the pure water layer to reduce the environmental background.

The light signals in the LS are detected by thirty PMTs mounted on the stainless steel truss between the acrylic vessel and the tank. All the PMTs are fixed on the truss through stainless steel brackets. The truss measures 1795 mm in diameter and 1820 mm in height. There is also a 10-mm-thick black acrylic shield between the inner acrylic vessel and the truss. It provides the holes to keep the PMT photocathode inside, thereby reducing the influence between internal and external light signals\linelabel{line:end-1ton-description}.

\begin{figure}[!htb]
    \centering
    \subcaptionbox{Acrylic vessel\label{fig:acrylic-vessel}}{\includegraphics[height=160pt]{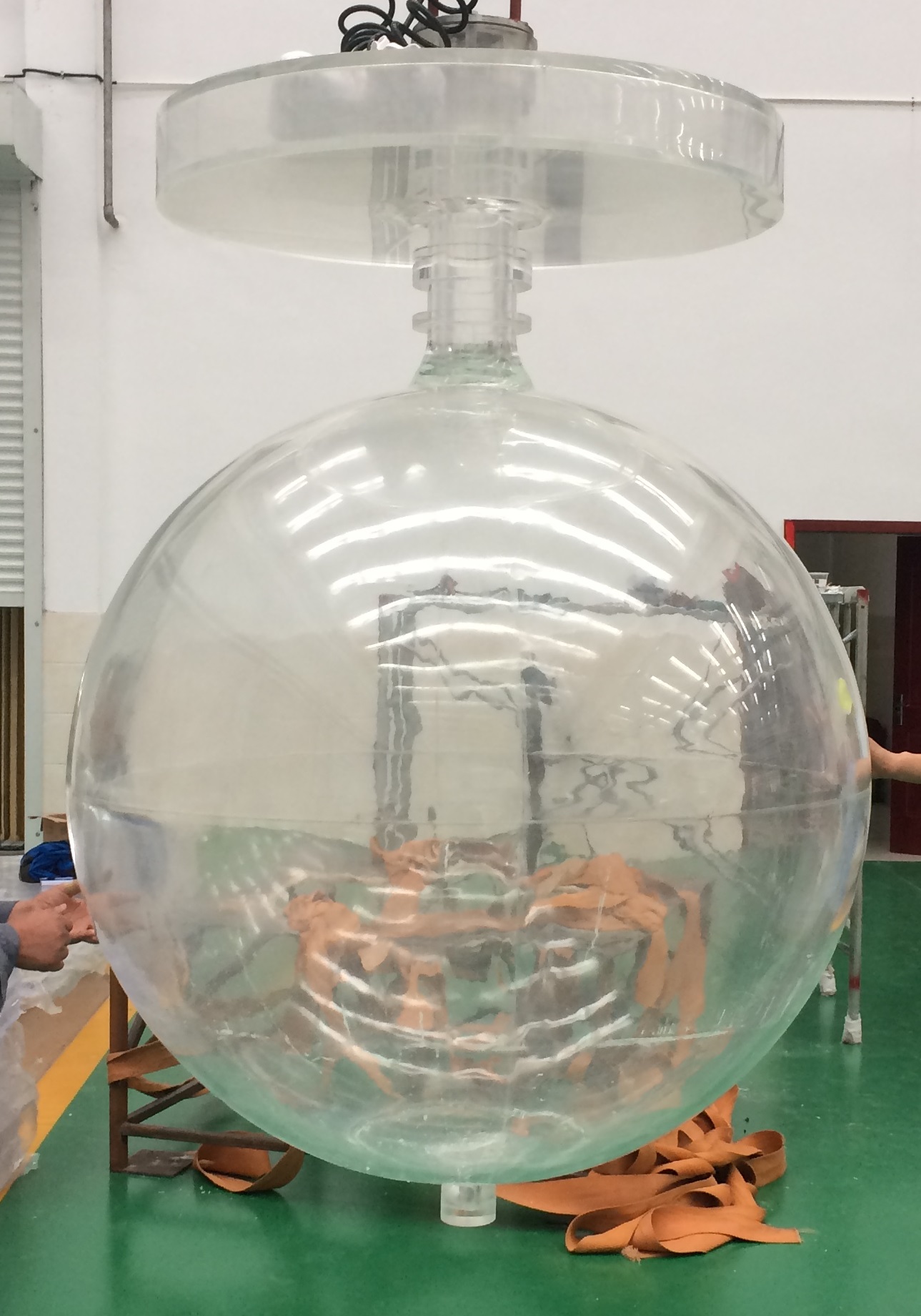}}
    \subcaptionbox{Inner detector\label{fig:core-detector}}{\includegraphics[height=160pt]{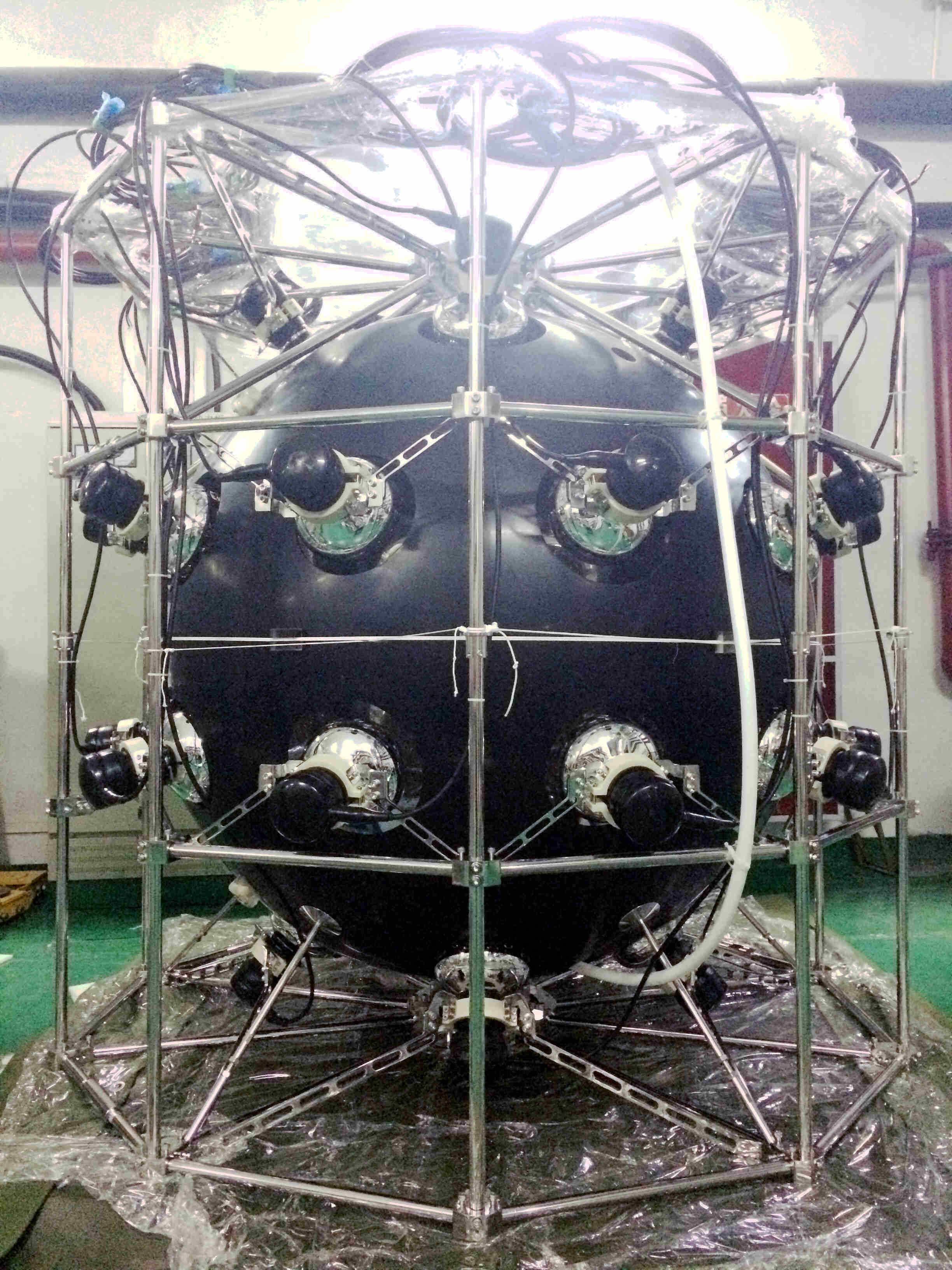}}
    \subcaptionbox{Water tank\label{fig:water-tank}}{\includegraphics[height=160pt]{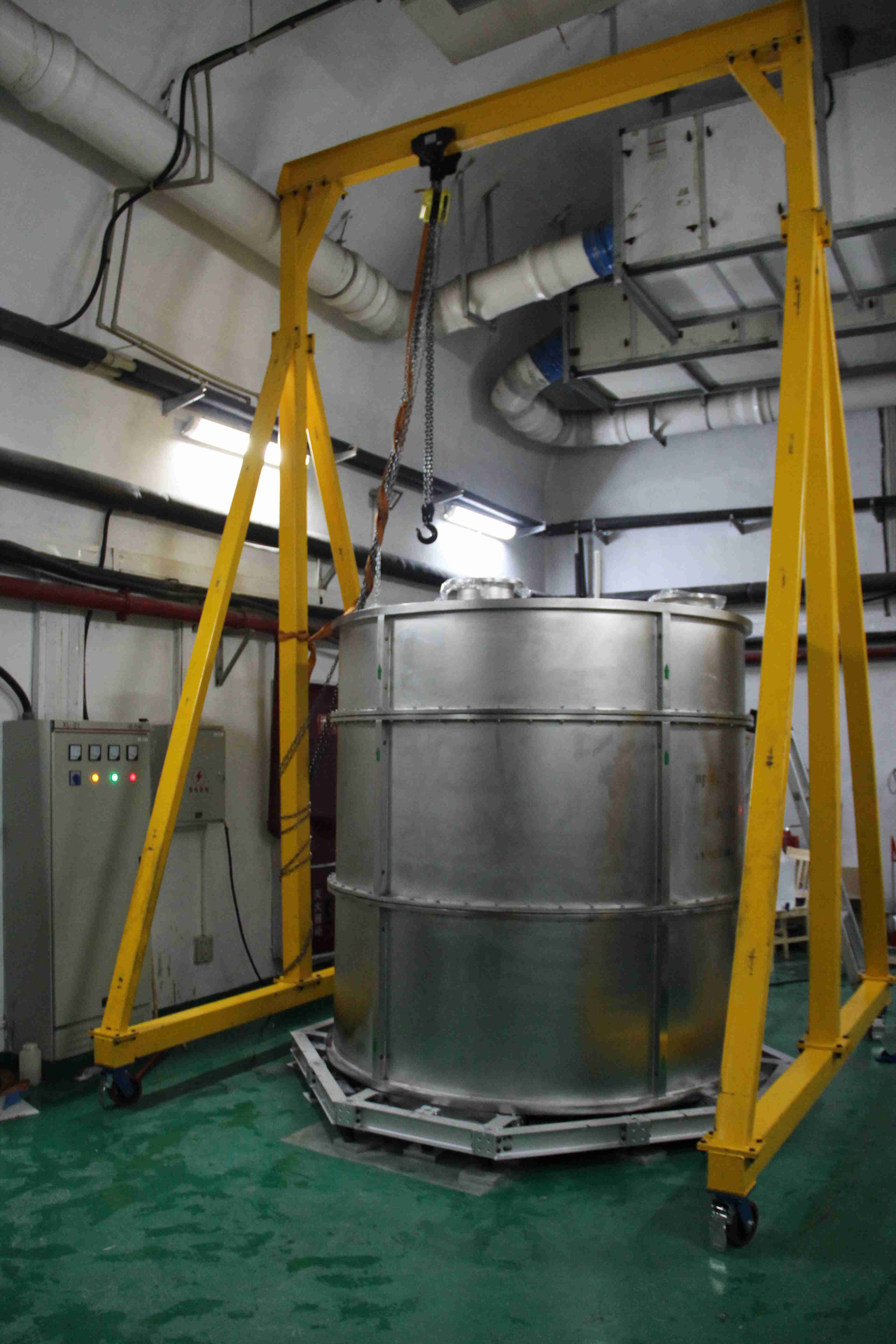}}
    \subcaptionbox{Overview from outside\label{fig:1ton-picture}}{\includegraphics[width=0.90\textwidth]{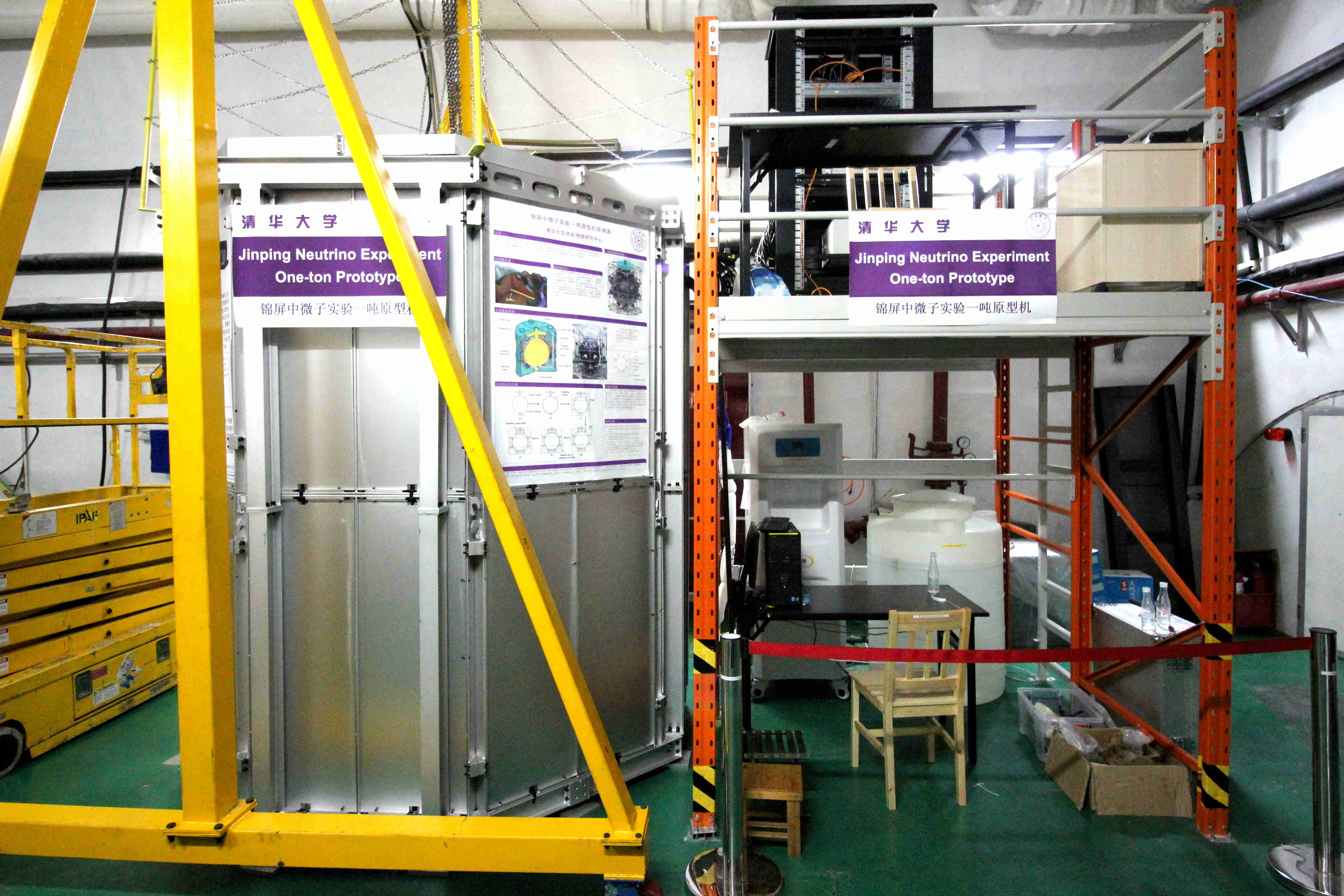}}
    \caption{The assembling of the 1-ton prototype at CJPL-I. Fig.~(a)-(c) show the part of the installation process of the prototype. Fig.~(d) shows the on-site workplace after the installation. In Fig.~(d), the pure water device and the water bucket are located on the right side of the 1-ton prototype. The electronics system is on the second floor of the shelf.}
    \label{fig:1tconstruction}
\end{figure}

\linelabel{line:1ton-installation}In 2017, the prototype was assembled at CJPL-I. Due to the limited operational space, the tank was built with three parts from top to bottom with corresponding heights of 470 mm, 800 mm, and 820 mm, respectively. The truss was split into ten sections along the perimeter and three parts from top to bottom. The corresponding heights of the three parts are 540 mm, 740 mm, and 540 mm, respectively. During the installation, the acrylic vessel was the first to be fixed and then mounted with the black shield. Then, the acrylic vessel and black shielding were installed onto the truss together and thirty PMTs were installed next. The whole inner structure was hoisted into the bottom part of the tank. Finally, the middle and top parts of the tank were also installed. Part of the installation process can be seen in Fig.~\ref{fig:1tconstruction}.

The design was confirmed with a detailed structural calculation to ensure its safety and reliability from the mechanical point of view. See more information in Ref.~\cite{Wang:2017ynm}.
\subsection{Detector components}

\subsubsection{Electronics}

The electronics system converts the light signals detected by PMTs into digitized waveforms, determines the global trigger, and records waveforms of triggered events to disk. Four CAEN V1751 flash analog-to-digital converter (FADC) boards, each with eight channels, are responsible for waveform digitizing. They are all configured as 10-bit ADC precision in \SI{1}{\volt} dynamic range and a \SI{1}{\giga\hertz} sampling rate. One logical trigger module, CAEN V1495, collects the over-threshold information of all channels from the FADC boards and determines the global trigger. We defined a qualified fired PMT to have its output voltage above \SI{10}{\milli\volt} before October 14, 2018, and \SI{5}{\milli\volt} in the later periods. A trigger signal was issued when over 25 PMTs got fired simultaneously in a \SI{125}{\nano\second} window before June 29, 2019. Later this threshold was lowered to 10 PMTs to reduce the impact of the total reflection that occurs on the inner surface of the acrylic vessel. After the trigger, digitized waveforms and event metadata of all channels are sent to the CAEN V2718 board via the VME bus, then forwarded to the CAEN A2818 PCIe card on data acquisition (DAQ) server via optical fiber with CAEN's proprietary protocol (CONET)\linelabel{line:DAQ protocol}. On the server, the readout program conducts a sanity check on the event timestamp and drops the under-threshold channels. Finally, waveforms and timestamps are written into the disk in ROOT format~\cite{ROOTarticle,ROOT}. The DAQ system recorded the PMT pulse shape with a \SI{1029}{\nano\second} time window for the fired PMTs initially. For experimental studies, the window length was later changed to \SI{600}{\nano\second}.

Table \ref{tab:tablePhase} lists the detailed operation conditions during different phases, P1, P2, P3, and P4. These phases cover all periods for which data are used in this study. The start and end dates are expressed in the format of Year-Month-Day. $T_{\mathrm{DAQ}}$ gives the DAQ live time of each phase. $N_{\mathrm{PMT}}$ gives the threshold of the number of fired PMTs. $T_{\mathrm{w}}$ gives the window length for recording a single triggered event. $H_{\mathrm{th}}$ represents the trigger thresholds for the single channel. The last column indicates whether the nitrogen bubbling system was installed and running during the corresponding phase.
\begin{table}[!htb]
    \caption{\label{tab:tablePhase}%
        The details of each phase for the 1-ton prototype. $T_{\mathrm{DAQ}}$: DAQ time. $N_{\mathrm{PMT}}$: trigger threshold (number of the fired PMTs). $T_{\mathrm{w}}$: length of the waveform in a single channel. $H_{\mathrm{th}}$: trigger threshold of a single channel. Nitrogen bubbling: presence of nitrogen bubbling system.
    }
    \centering
    \resizebox{\textwidth}{12mm}{
        \begin{tabular}{cccccccc}
            \midrule
            Phase                                  &
            \textrm{Start date}                    &
            \textrm{End date}                      &
            $T_{\mathrm{DAQ}}$ [day]               &
            $N_{\mathrm{PMT}}$                     &
            $T_{\mathrm{w}}$ [\unit{\nano\second}] &
            $H_{\mathrm{th}}$ [mV]                 &
            Nitrogen bubbling                                                                                \\
            \midrule
            P1                                     & 2017-07-31 & 2018-10-14 & 392.02 & 25 & 1029 & 10 & No  \\
            P2                                     & 2018-10-15 & 2019-06-29 & 238.45 & 25 & 1029 & 10 & No  \\
            P3                                     & 2019-06-30 & 2019-07-14 & 12.60  & 10 & 600  & 5  & No  \\
            P4                                     & 2019-07-15 & 2020-09-27 & 177.21 & 10 & 600  & 5  & Yes \\
            \midrule
        \end{tabular}
    }
\end{table}

\subsubsection{Liquids}
With a reasonable light yield and time constants, linear alkylbenzene (LAB) dissolved with \SI{0.07}{\gram/\liter} 2,5-diphenyl\-oxazole (PPO) and \SI{13}{\gram/\liter} of 1,4-bis (2-methyl\-styryl)\linebreak-benzene (bis-MSB) has been used as a slow LS~\cite{Guo:2017nnr} and implemented in the 1-ton prototype. This slow LS candidate provides the capability to separate the \v{C}herenkov light from the scintillation light with a relatively high light yield, i.e., \SI{4010}{photons\per\MeV}. More details on the LS experimental study can be found in Ref.~\cite{Li:2015phc,Guo:2017nnr}.

\begin{figure}[!htp]
    \centering
    \label{fig:purewatercirculation}
    \includegraphics[width=0.6\linewidth]{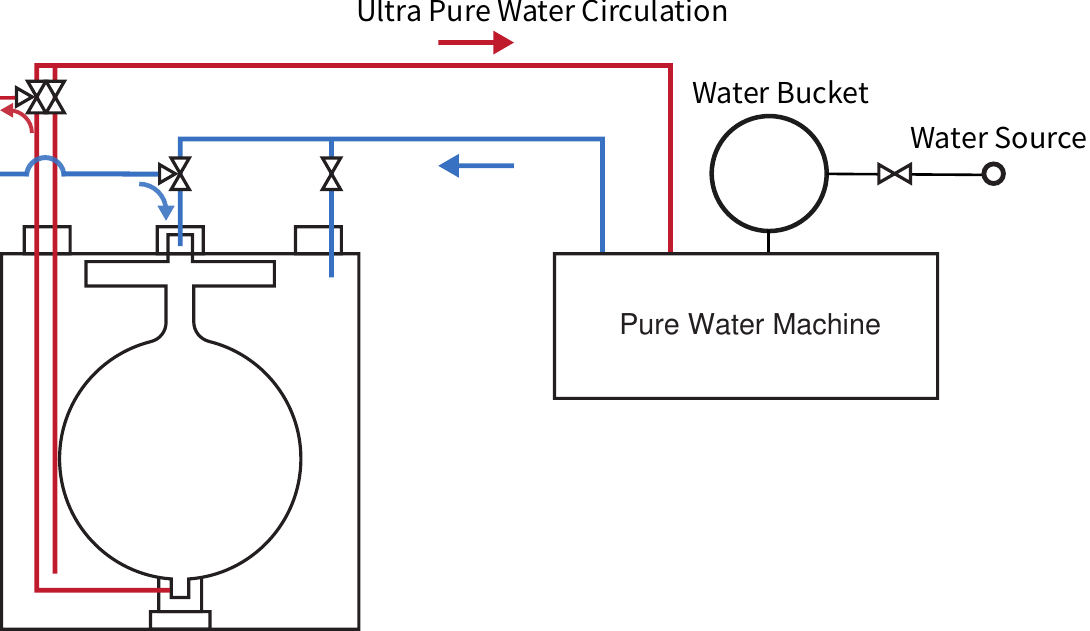}
    \caption{The pure water circulation system for the 1-ton prototype. Red and blue lines represent pipes for the water coming from the prototype, and the injecting ultrapure water, respectively. Before P1, there was a short period when the central acrylic vessel filled with water. At that time, two 3-way valves were configured to connect to the pure water device, forming a cycle for the inner detector. After that, they were turned to outward connections, for LS filling and the \ce{N2} gas system.}
    \label{fig:PureWaterSystem}
\end{figure}

\linelabel{line:water system}A pure water system is used to purify the water in the stainless steel tank.
Before P1, there was a short period when the central acrylic vessel filled with water.
The water system consists of an all-in-one electrodeionization (EDI) pure water device and a \SI{500}{\liter} bucket, which can be seen in Fig.~\ref{fig:1tconstruction} (d). The model of the pure water device is ZYpure-EDIc-100-UP~\cite{ZYPURE}. Fig.~\ref{fig:PureWaterSystem} shows the circulation system. The raw water supplied by the laboratory is first cached in the \SI{500}{\liter} bucket, and purified through multiple preprocess filters, reverse osmosis, and EDI. Then the post-processing system, including an ion-exchange resin, further filters the water to about \SI{18}{\mega\ohm\centi\meter} and pumps it to the tank at about \SI{100}{\liter} per hour. Water coming out from the tank will again enter the post-process system and get recycled. \linelabel{line:sensors}Two Suntex EC-4100-RS~\cite{EC-4100-RS} resistivity sensors were deployed to monitor the water resistance and temperature. Also, two Senix TSPC-30S1-232~\cite{TSPC-30S1-232} ultrasonic sensors were installed to monitor the levels of liquids in the tank and the acrylic vessel. The ultrasonic sensors were turned off during DAQ to avoid electromagnetic interference on readout electronics.

\subsubsection{PMTs}
A total of thirty \linelabel{line:pmt-model} Hamamatsu\cite{Hamamatsu} R5912 eight-inch PMTs are installed in the detector, with a photocathode coverage of about 12\%.
Fifteen of them use negative high voltage (HV) and are manufactured with low background glass. The remaining fifteen PMTs use positive HV and normal glass material.
They are alternately installed in four layers. Five PMTs are in the top and bottom layers, and ten PMTs in the middle two layers. The HVs of the thirty PMTs are from \SI{1600}{\volt} to \SI{1700}{\volt} for gain alignment. In this HV range, the transit time (TT) and transit time spread (TTS, FWHM) are approximately \SI{55}{\nano\second} and \SI{2.4}{\nano\second}, respectively. The typical rise time of waveforms is \SI{3.8}{\nano\second}. The quantum efficiency reaches a maximum of 25\% at about \SI{390}{\nano\meter}.

\linelabel{line:HV system}Wiener Mpod HV crate with Iseg EDS 30330p and 30330n HV modules~\cite{WIENERHV} provide HV for all PMTs. A computer controls the HV of each channel via SNMP protocol.

\subsubsection{Gas system}\label{sec:gas-system}
\linelabel{line:Nitrogen system}No gas system was deployed with the initial detector installation. The current nitrogen gas system was installed between July 11 and 14, 2019. A nitrogen purging and sealing experiment was proposed within the collaboration due to the radon leakage indicated by the $\beta$-$\alpha$ cascade decay of \ce{^{214}Bi}-\ce{^{214}Po}, which will be detailed in Sec.~\ref{sec:BiPo214}. The radon leakage hinted that the detector was not airtight enough, which causes radon entering the detector along with air from the experimental hall. Preliminary studies also suggested that oxygen quenching effect weakens the detector's ability to discriminate $\beta$/$\alpha$ particles.

For these reasons, we installed a nitrogen bubbling system that provides a positive pressure inside the detector, \linelabel{line:fix-airtightness1}to avoid air leaking into the acrylic vessel. At the same time, the removal of the oxygen dissolved in the LS was achieved in the degassing process with nitrogen, mitigating the quenching effect. As a consequence, the increase in light yield has been observed in P4 as expected.

\begin{figure}[!htb]
    \centering
    \includegraphics[width=0.6\linewidth]{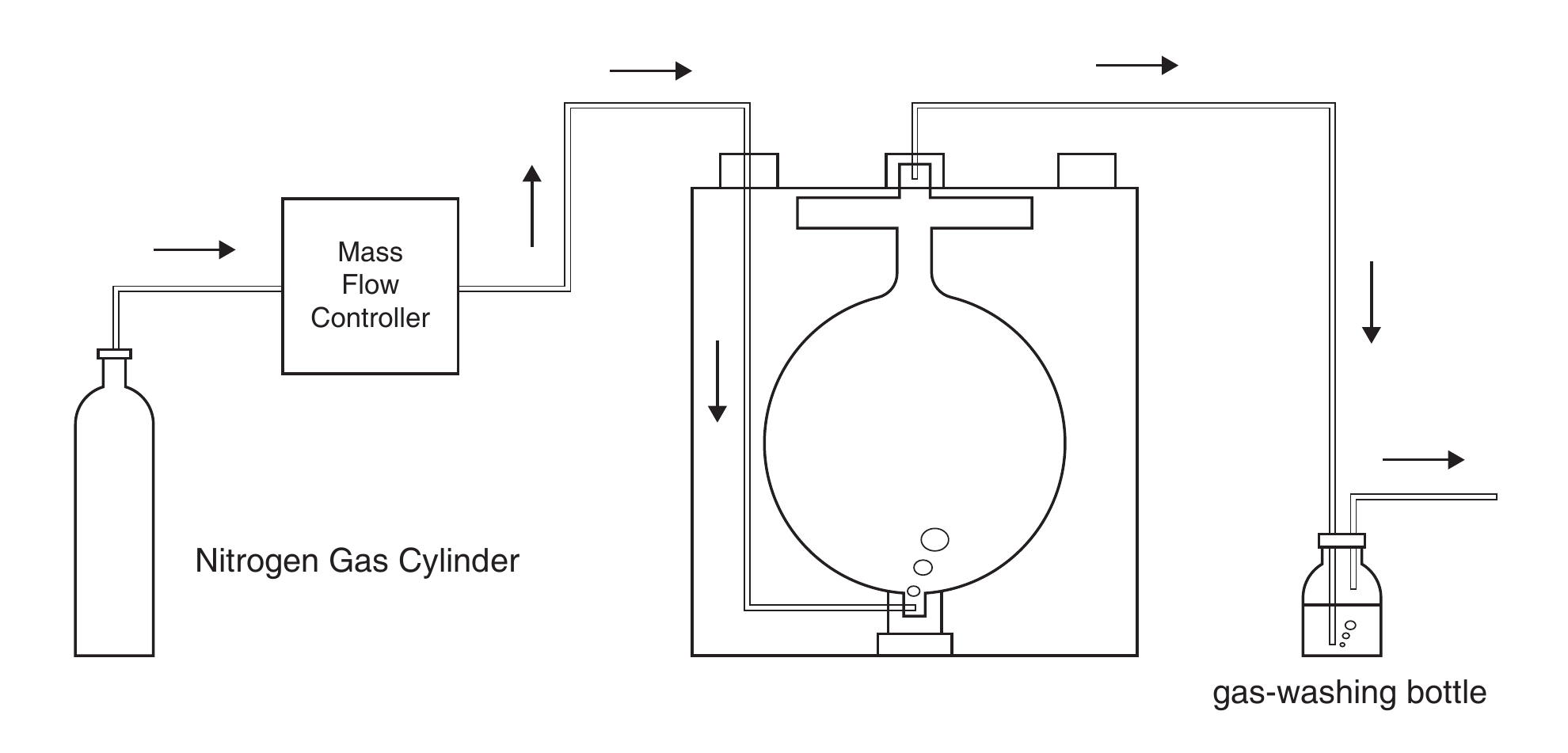}
    \caption{The nitrogen bubbling system of the 1-ton prototype. Nitrogen gas comes from the leftmost cylinder, then passes through the gas mass flow controller. Then the gas is released from the bottom of the LS ball and finally gets discharged through the washing bottle.}
    \label{fig:Nitrogen system}
\end{figure}

As shown in Fig.~\ref{fig:Nitrogen system}, nitrogen gas comes from a high-pressure cylinder of liquid nitrogen. A relief valve is installed on the cylinder to adjust the nitrogen pressure. A gas mass flow controller SLD-MFC 600D~\cite{SLD-MFC-600D} controls the gas flow to be approximately \SI{10}{\milli\liter/\minute}. The gas then goes into the bottom of the LS and out from a gas washing bottle. \linelabel{line:fix-airtightness2}This process is checked by counting the bubble rate in the washing bottle.

\subsection{Data quality}
\label{sec:DataQuality}

Before the detector calibration in Sec.~\ref{subsec:detector-cali} and background analysis in Sec.~\ref{sec:measurement}, about 0.64\% of the data were excluded because of bad quality. The basic strategy of data quality check is to analyze the collected data offline to find out abnormal DAQ periods. We exclude anomalous data periods by investigating various properties, including the overall event rate, as well as trigger rate and waveform baseline of each PMT. The time interval for data quality check is about 5 minutes, while the exact length depends on the event rate.

There are two methods to determine anomalies in data quality check. The first one compares data quality indices, e.g. the average trigger rate of its nearby periods. We treat the distribution of the data quality index as Gaussian, and if the index of the period is 5$\sigma$ away from its nearby average, this period is marked as abnormal. The event rate is a global quality index, which indicates the overall quality of the whole detector. Other indices, including the occupancy or baseline, are specific to individual channels. When an index goes abnormal, only the corresponding channel is tagged as bad and excluded from subsequent calibrations.

The second method is checking baseline stability. The baseline and its standard deviations of all readout waveforms are calculated offline. If a channel had extremely high baseline fluctuation in a period, then it was possibly suffered from large electronic noises or other detector malfunctions. That channel in the period will be marked as bad. When the number of bad channels in one period exceeds 15, the data of that period, which includes all channels, will be marked as abnormal.

\subsection{Charge-weighted vertex reconstruction}
The reconstructed vertex $\hat{\mathbf{r}}$ is estimated by the charge-weighted average of PMT positions,
\begin{equation}\label{eq:vertex-recon}
    \hat{\mathbf{r}}=c \cdot \frac{\sum_{i=1}^N q_i \mathbf{r}_i}{\sum_{i=1}^N q_i},
\end{equation}
where $q_i$ is the number of PEs collected by the $i$th PMT, $\mathbf{r}_i$ is the position of the $i$th PMT, and $c$ is a correction factor. The value of $c$ depends on the detector geometry. This method can be found in spherical scintillator detectors for fast vertex estimation~\cite{Liu_2018}. Since the detector scale is much smaller than the attenuation length ($\sim$10 m)~\cite{Goett:2011zz,xiao2010study}, the optical attenuation effect is negligible in our reconstruction.

The correction factor $c$ is determined to be 1.5 theoretically.
\begin{figure}[!htb]
    \centering
    {\includegraphics[width=0.6\linewidth]{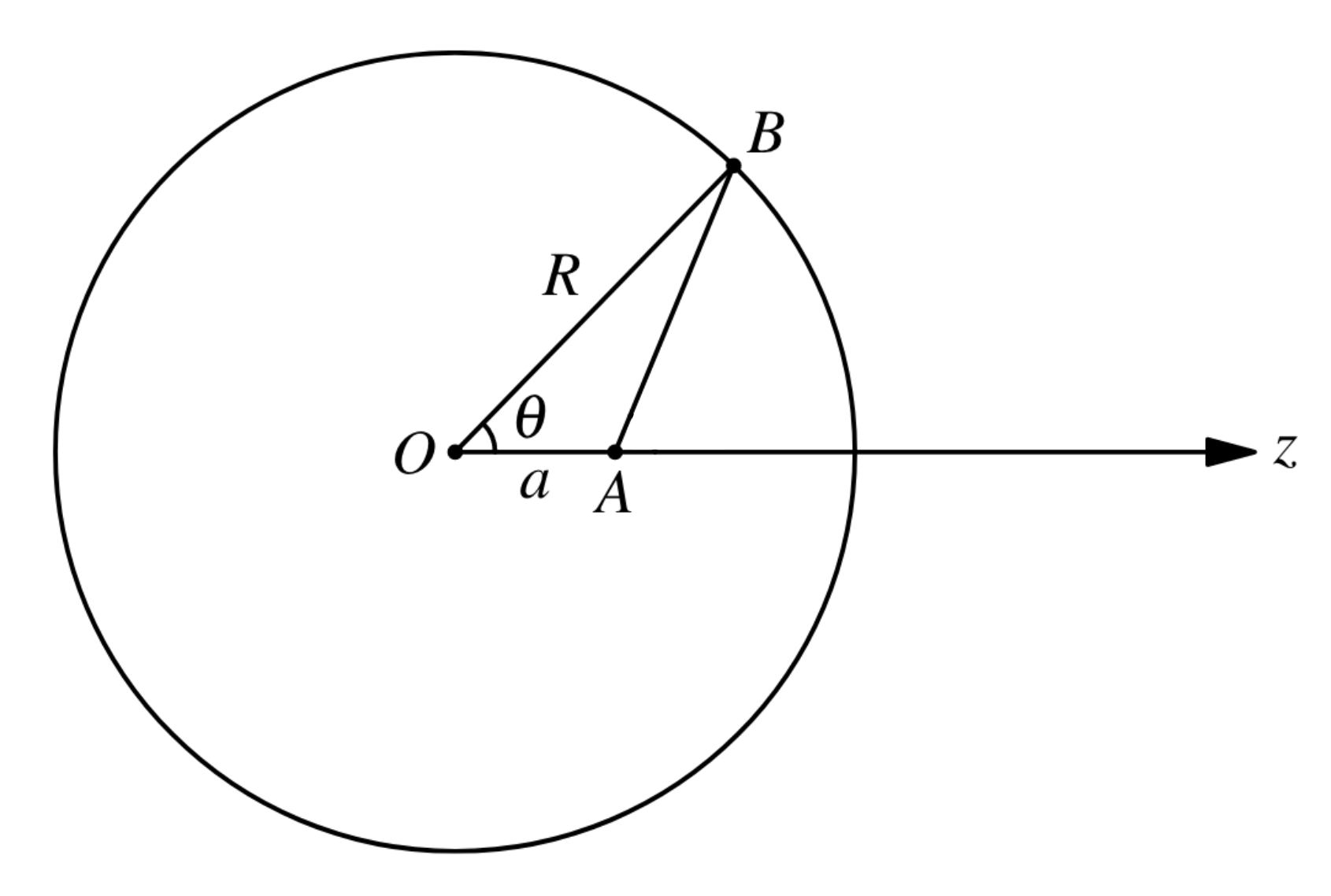}}
    \caption{An ideal spherical detector. The vertex is placed at $A$. One of the PMTs is placed at $B$.}
    \label{fig:SphereShow}
\end{figure}
As shown in Fig.~\ref{fig:SphereShow}, taking the 1-ton prototype as an ideal spherical detector and assuming that point $A$ is placed at a distance $a$ from the spherical center $O$, the PE density $\sigma$ received by a PMT at the point $B$ on the sphere from $A$ can be predicted,
\begin{equation}
    \sigma=Q \cdot \frac{\mathrm{d} \Omega}{4 \pi} \frac{1}{\mathrm{~d} S}=\frac{Q}{4 \pi} \cdot \frac{R-a \cos \theta}{\left(a^2+R^2-2 a R \cos \theta\right)^{3 / 2}},
\end{equation}
where $R$ is the radius of the sphere, $\theta$ is the angle between $\overrightarrow{OA}$ and $\overrightarrow{OB}$, $Q$ is the total number of the PE emitted by one event, $\mathrm{d}S$ is the area of the PMT photocathode, and $\mathrm{d}\Omega$ is the solid angle formed by $\mathrm{d}S$ to the spherical center $O$. The estimated vertex $\hat{\mathbf{r}}$ in Eq.~\ref{eq:vertex-recon} can be re-written in an integral form,
\begin{equation}
    \hat{\mathbf{r}}=\frac{c}{Q} \int \mathbf{r} \sigma \mathrm{d} S.
\end{equation}
Carry out the integral, it gives that $\hat{r}=2 c a/3$. Setting the normalization factor $c=1.5$ makes the reconstructed radius $\hat{r}$ equal to the actual radius $a$.

\begin{figure}[!htb]
    \centering
    {\includegraphics[width=0.6\linewidth]{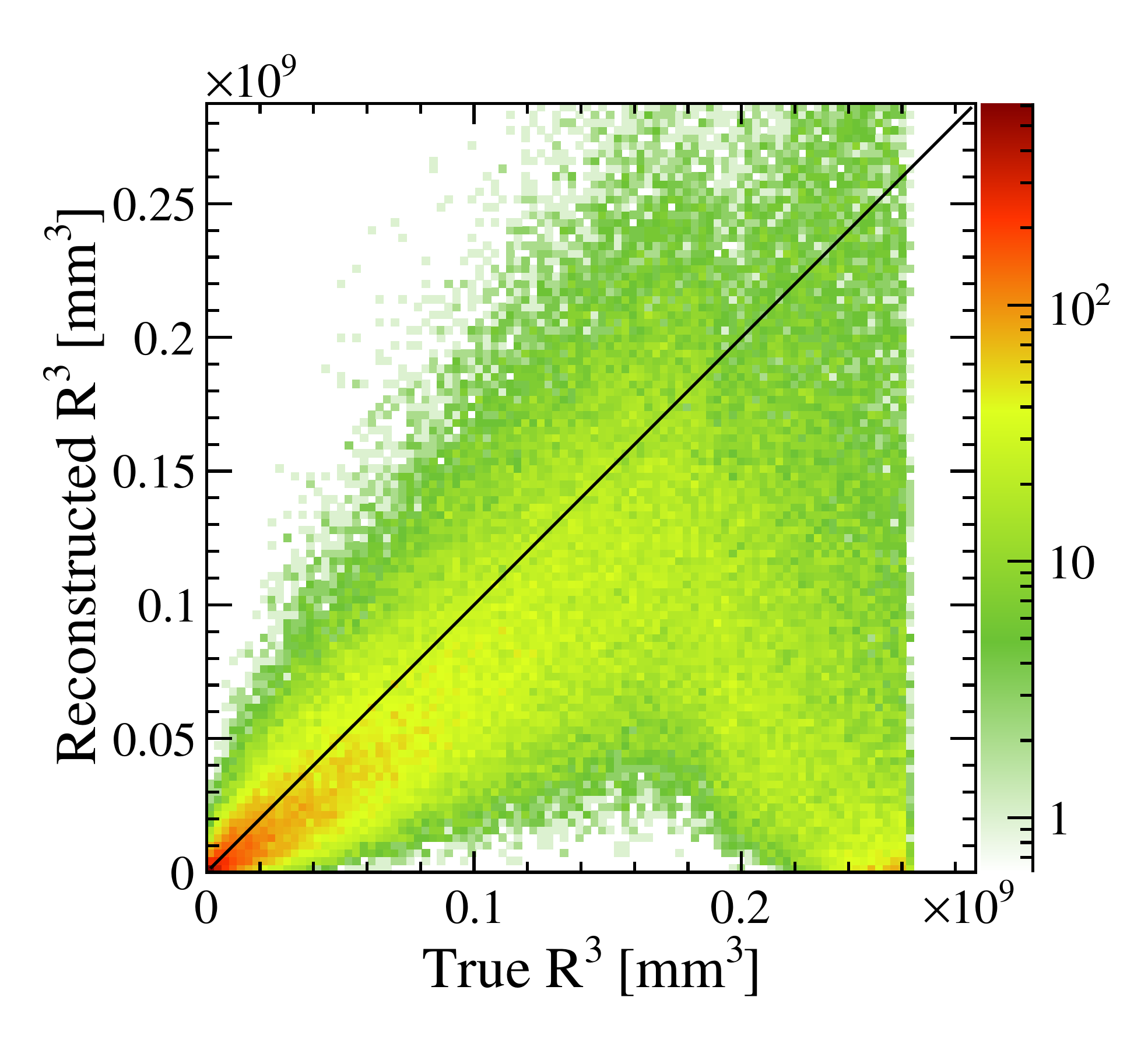}}
    \caption{The reconstructed $R^3$ versus actual $R^3$ for 1-MeV electron generated uniformly in the LS of the 1-ton prototype in the MC simulation. The black line represents the case that reconstructed $R^3$ equals true $R^3$.}
    \label{fig:vertex}
\end{figure}

To examine the effect of the vertex reconstruction algorithm, Fig.~\ref{fig:vertex} shows the actual radius versus the reconstructed radius for the Monte Carlo (MC) sample. The resolution of the vertex reconstruction using Eq.~\ref{eq:vertex-recon} is evaluated to be about \SI{11.5}{cm}\linelabel{line:vertex resolution}. It can be seen from Fig.~\ref{fig:vertex} that the reconstructed vertices in the region close to the detector center are scattered near the actual vertex positions, as expected. While at the detector boundary, due to the significant total reflection effect, the reconstructed vertex positions show a systematic bias from the truth.
Studies~\cite{Dou:2022glt,Luo:2022xrd} of spherical detectors deployed with more PMTs in the simulation shows that the reconstruction algorithm can significantly improve with more coverage of PMTs.

\linelabel{line:beginTotalReflection}
The total reflection effect at the boundary of the acrylic sphere and water in tank has been studied via both MC simulation and theoretical calculation. Replacing the water in the tank with a material with the same refractive index as acrylic, e.g.~mineral oil, can avoid this effect.

The photons generated in LS need to pass through the acrylic container layer to hit each PMT. The refractive index of acrylic is greater than that of water. In this case, some photons encounter total reflection and thus can't be detected. A survey of the refractive index of acrylic and water indicates that the total reflection angle is about $[54^\circ$,~$64^\circ]$ for wavelength in $[200,~800]$ nm.

As shown in Fig.~\ref{fig:SphereShow}, in the spherical coordinate system, the incident angle $\alpha$ of the luminous point $A$ on the spherical surface can be expressed as,
\begin{equation}\label{eq:total-reflection-angle}
    \alpha\left(\theta, d\right)=\arcsin \left(\frac{d\sin \theta}{\sqrt{d^2+1-2d \cos \theta}}\right),
\end{equation}
where $\theta$ is the zenith angle as shown in Fig.~\ref{fig:SphereShow}, $d$ is defined as $a/R$. $\alpha$ first increases and then decreases with $\theta$ increasing from $0^\circ$ to $180^\circ$. Therefore, $\alpha$ reaches the maximum value ($\alpha_{\mathrm{max}}$) for a medium size of $\theta$, which is about $15^\circ\sim30^\circ$ in our case. It leads to the fact that, in spherical detectors, the region of the incident point on the spherical surface where total reflection can occur is a ring belt centered on the $z$-axis shown in Fig.~\ref{fig:SphereShow}. The PMTs in this area receive much less charge than other PMTs. $\alpha_{\mathrm{max}}$ always increases when $d$ increases and can exceed 54$^\circ$ when $d>0.8$, indicating that the total reflection gradually becomes apparent for events in the region further than $0.8R$ away from the detector center. For our 1-ton prototype, this distance threshold is about 0.52 m.
\linelabel{line:stopTotalReflection}
\subsection{Detector calibration}
\label{subsec:detector-cali}
The calibrations of the 1-ton prototype include the PMT gain calibration, PMT time calibration, and energy scale calibration. The calibration parameters are also indicators of the running status.

\begin{figure}
    \centering
    \resizebox{0.8\textwidth}{!}{
        \begin{tikzpicture}[node distance=5em, text width=5em]
            \node[startstop](FADC){FADC\\Waveform};
            \node[startstop, right of = FADC, xshift=3em](Charge){Charge \\(ADC$\times$ns)};
            \node[startstop, right of = Charge, xshift = 3em](NPE){$\mathrm{N_{PE}}$};
            \node[startstop, right of = NPE,xshift = 3em](Energy){Energy (MeV)};
            \node[startstop, below of = NPE](Vertex){Vertex};
            \node[startstop, below of = Energy](Time){Time};
            \draw [arrow] (FADC)  -- node [above, yshift=1.3em] {Integration} (Charge);
            \draw [arrow] (Charge)  -- node [above, yshift=1.3em, text width=5em,align=center] {Gain\\Calibration} (NPE);
            \draw [arrow] (NPE)  -- node [above, yshift=1.3em, text width=8em,align=center] {Energy Scale Calibration} (Energy);
            \draw [arrow] (NPE)  -- node [left, text width=7em, yshift=-0.5em, align=center] {Vertex\\Reconstruction} (Vertex);
            \draw [arrow] (Vertex)  -- node [above, yshift=0.65em, text width=5em,align=center] {Time\\Calibration} (Time);
        \end{tikzpicture}
    }
    \caption{Flow chart of the calibration and reconstruction. The details of the calibration and vertex reconstruction processes are introduced in this section.}
    \label{fig:CalibrationProce}
\end{figure}
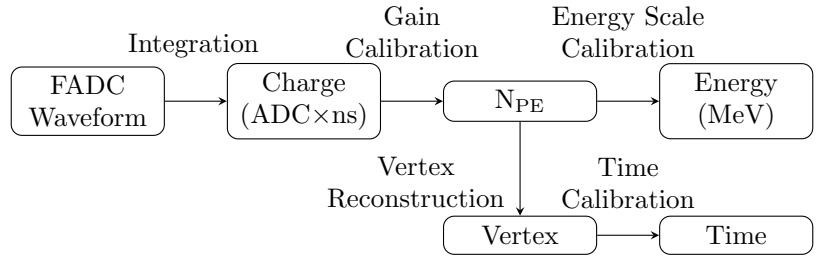

Our calibration approach is based on the dark noise and internal radioisotope decay products, such as $\alpha$'s and $\gamma$'s. Fig.~\ref{fig:CalibrationProce} shows several key steps in the calibration processes. Starting from the original FADC waveform, the PMT charge was obtained by an integration and was used in the gain calibration. Then, by dividing the calibrated gain, the charge unit was converted to the number of PE ($N_{\mathrm{PE}}$). Afterward, using the $N_{\mathrm{PE}}$ of each PMT, the vertex was reconstructed using Eq.~\ref{eq:vertex-recon}. Reconstructed events near the detector center were used for time calibration. Meanwhile, the energy scale (PE/MeV) calibration was performed by locating the monoenergetic $\gamma$ peak from \ce{^{208}Tl}. The $\gamma$ from \ce{^{40}K} and $\alpha$ from \ce{^{214}Po} decay provided the cross-checks for the energy scale calibration. The following sections detail these steps.

\subsubsection{Gain}
We used a ``RollingGain'' method to perform a gain calibration. This method utilizes the dark noise in each physics run, so that the gain parameters are updated on a rolling basis for each run. The gain factor also characterizes the health of a PMT. It is used for online data quality monitor and offline data processing.

The calibration is to obtain the charge distribution of a PMT and fit the gain factor with a probability distribution function deduced from the PMT response model~\cite{Bellamy:1994bv}. The charge ($Q$) distribution of signals from a PMT without backgrounds was modeled using the convolution of a Poisson distribution with a Gaussian function.

The main component of dark noise is the signal formed by photocathode thermal emission electrons. It is essentially the same as the signal formed by a single PE. The dark noise in data is required to appear before the signal region, i.e., within the first \SI[parse-numbers=false]{150~(40)}{\nano\second} for \SI[parse-numbers=false]{1029~(600)}{\nano\second} time window size. The peak height is required to be \SI{>4}{\milli\volt} to avoid baseline fluctuation. The time to the previous trigger is required to \SI{>20}{\micro\second} to suppress the effect of re-triggering and ringing. Only one PMT is allowed to have dark noise hit in an event to avoid the physical signal appearing in the dark noise window. We integrate the waveform within [-20,~+30] \unit{\nano\second} around the peak time of the selected dark noise to calculate its charge. The derived charge distribution is then collected every two weeks to provide a fitting result with a precision of about 1\%.

\linelabel{line:gain-bkg-begin}
Two kinds of backgrounds are considered for the dark noises. One is induced by the low charge signals, including white noise present in each event. It makes the width of the signal distribution non-zero when there is no {PE} emission from the photocathode. A Gaussian distribution was used to describe this background. The other is due to the discrete processes accompanying the measured signal with non-zero probability, such as thermal emission from other dynodes and noise induced by the measured light. Exponential functions were used to describe this background.
\linelabel{line:gain-bkg-end}

The measured dark noise rate of the negative and positive HV PMTs are about \SI{3500}{\hertz} and \SI{500}{\hertz}, respectively. The former is an order of magnitude higher than the latter because the voltage of the negative HV PMTs is applied to the photocathode. Besides the traditional dark noise from thermal electrons directly amplified by dynodes, those electrons from photocathode may also hit the glass bulb and produce glass scintillation, since its surrounding water layer and steel barrel are connected to the ground.

\subsubsection{Time}
In the 1-ton prototype, photons simultaneously hit on each PMT may have different readout times due to: different TT and TTS for each PMT; different cable lengths of the PMTs; asynchronous time between the four boards for phase P1 and P2; time offsets between different channels of the same FADC board.

The time calibration aims to minimize the above effects on the reconstructed event time. It allows the direction reconstruction of muons using hit times on each PMT. In this study, $\mathrm{TT}_j$ and $\mathrm{TTS}_j$ include all the above effects that may introduce time offsets between channels.

Different from other neutrino experiments~\cite{liu2014automated,back2012borexino,moffat2005optical,abe2014calibration}, we developed a method using natural radioactive events for time calibration.

In a conventional TDC circuit, once a PMT pulse crosses the threshold, a timestamp is given and recorded as an edge by TDC. The time of the edge defines the arrival time of the pulse. However, the rising edge of a larger pulse can cross the thresholds earlier, yielding an earlier arrival time, which is called the time slewing effect~\cite{xuzong2007time}. In the 1-ton prototype, the threshold is set to dynamic, i.e., 10\% of the first peak in the signal region, to avoid this effect.

Performing a time calibration using low-energy events may suffer from long scintillation decay times, especially as the target material for the 1-ton prototype is slow LS. Therefore, we select events with large PE number on at least two PMTs and require these events to be reconstructed near the detector center. The time of flight (TOF) uncertainty due to poor vertex resolution on detector edge can also be suppressed in this way.

For each event $i$, we define $t_i$ as its beginning time. The TOF of photon from the reconstructed vertex to the fired PMT $j$ is denoted as $\mathrm{TOF}_{ij}$, so the photon hit time is $t_i + \mathrm{TOF}_{ij}$. The rise time of the first PE pulse relevant to this event for PMT $j$ is $\tau_{ij}$, which can be directly acquired via offline waveform processing. The time shift between PE hit time and pulse rise time is $T_j$, shared by all physical events, which is the goal of time calibration. The relationship between these variables can be expressed as
\begin{equation}
    \tau_{ij} = t_i + T_j + \mathrm{TOF}_{ij}.
    \label{eq:timecalibeq}
\end{equation}

Shifting $T_j$ to $T_j + \Delta T$ for all PMT does not bring any observable effects. Thus, we chose the gauge $\sum_{j\in\{\text{PMT Id}\}} T_j = 0$. Due to TTS and other uncertainties, $T_j$ is a random variable following the normal distribution $\displaystyle N\left(\mu_j=\mathrm{TT}_j, \sigma_j^2\right)$ where $\sigma_j \propto \mathrm{TTS}_j$.

$\mathrm{TT}_j$ and $\mathrm{TTS}_j$ were determined by obtaining $t_i$ and $T_j$ in Eq.~\ref{eq:timecalibeq} with the least square method. An iterative algorithm was performed to minimize $\sum_{i,j} (\tau_{ij} - t_i - T_j - \mathrm{TOF}_{ij})^2$, by adjusting $t_i$ and $T_j$ alternatively. After the iteration converges, we obtained $T_j$ from the final step. $\mathrm{TTS}_j$ is the FWHM of the residue distribution $\tau_{ij} - t_i - T_j - \mathrm{TOF}_{ij}$.

To cross-check the time calibration, three independent analyses were performed, and they provided consistent calibration results. Based on the calibration result, we observed improvements in muon direction reconstruction, which verified the correctness of our calibration. %

\subsubsection{Energy scale}
\label{sec:EScale}
The task of the energy calibration is to convert the detected PE number of the event to the deposited energy in the unit of MeV. In the 1-ton prototype, we did not deploy an artificial calibration source, and completed the energy calibration by using the natural decay signals generated within the detector materials, namely, 1) $\gamma$ signal from \ce{^{208}Tl} decay (\ce{^{208}Tl}-$\gamma$); 2) $\gamma$ signal from \ce{^{40}K} decay (\ce{^{40}K}-$\gamma$); 3) $\alpha$ signal from \ce{^{214}Po} decay (\ce{^{214}Po}-$\alpha$).

We followed the multiplicity selection method described in Ref.~\cite{Yu:2013cob} to process events. This method considers the time correlation between events and divides the events into isolated events, double coincidences, triple coincidences and so on. This method is widely used in reactor neutrino experiments\cite{DayaBay:2014fud,RENO:2019otc} to find the inverse $\beta$-decay signals. The detailed event selection process is organized in Sec.~\ref{sec:measurement}.

\begin{figure*}[!htb]
    \centering
    \includegraphics[width=0.98\linewidth]{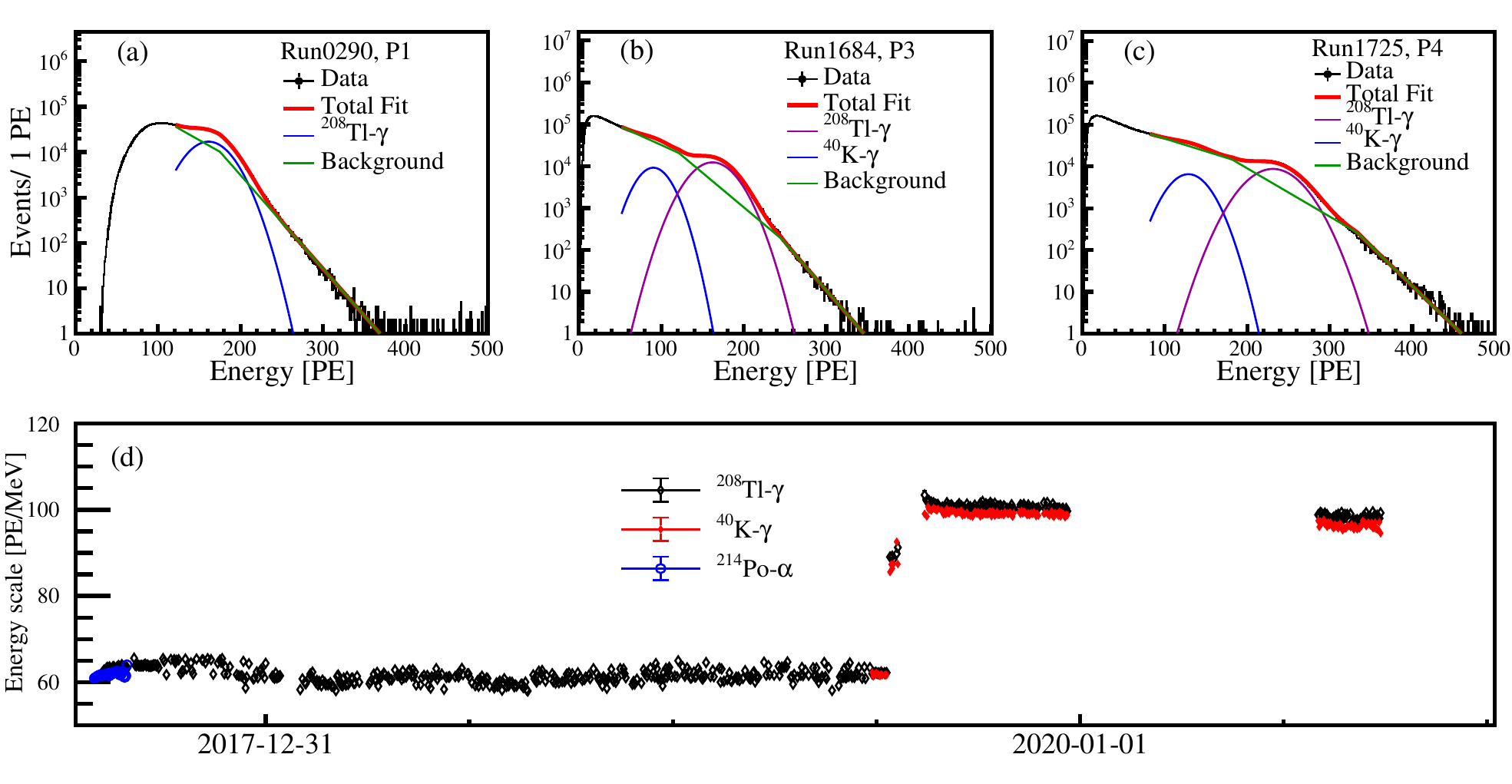}
    \caption{Fig.~(a), (b) and (c) show the typical energy spectrum of the isolated events during P1/P2, P3 and P4, respectively. The red lines represent the total fit results, namely, signal(s) plus backgrounds. While the pink and blue lines represent the fit results of \ce{^{208}Tl}-$\gamma$ and \ce{^{40}K}-$\gamma$, respectively. Fig.~(d) shows the fitted energy scale (\unit{PE/\MeV}) for three signals versus the DAQ date.}
    \label{fig:SpectrumFit}
\end{figure*}

After the multiplicity selection, we focused on the isolated events and double coincidences in the subsequent analysis. Most of the isolated events are the $\gamma$ and $\beta$ backgrounds from the natural decay in the detector material and environment. The double coincidences may contain the prompt and delayed signals formed by the cascade decay, and the other is mainly the random combination of the uncorrelated single events. Because the detector running conditions changed several times, the energy spectrum of the isolated events varies accordingly. Fig.~\ref{fig:SpectrumFit} (a)-(c) shows the typical energy spectrum of the isolated events in stages P1/P2, P3, and P4, respectively. The difference between Fig.~\ref{fig:SpectrumFit} (a) and (b) is due to the change of trigger threshold on June 30, 2019, from 25 fired PMTs to 10 fired PMTs, resulting in a significant increase of the low-energy event counts. The difference between Fig.~\ref{fig:SpectrumFit} (b) and (c) is due to the installation of a nitrogen bubbling system for the prototype in July 2019. \linelabel{line:fix-airtightness3}It reduced the effect of oxygen quenching on the energy deposition in LS, thus increased the overall energy scale.

The highest energy peak in Fig.~\ref{fig:SpectrumFit} corresponds to the \SI{2.61}{\MeV} $\gamma$ from \ce{^{208}Tl} decay. While the peak around 100 PE in Fig.~\ref{fig:SpectrumFit} (b) is the $\gamma$ of \SI{1.46}{\MeV} from \ce{^{40}K} decay. To fit the $\gamma$ peaks formed by the decay of \ce{^{208}Tl} and \ce{^{40}K} from Fig.~\ref{fig:SpectrumFit} (a)-(c), we first investigated the \ce{^{208}Tl} peak individually due to its better signal-to-background ratio. We used Gaussian to describe the $\gamma$ peak, and used the exponential functions to describe the backgrounds. Since we could not analyze the exact source of the backgrounds, we chose a variety of functions and fit conditions to validate the fit result. In Fig.~\ref{fig:SpectrumFit} (a)-(c), the red lines represent the total fit result, namely, signal(s) plus backgrounds. While the pink and blue lines represent the fit results of \ce{^{208}Tl}-$\gamma$ and \ce{^{40}K}-$\gamma$ respectively. These fitted energy scales (\unit{PE/\MeV}) are shown in Fig.~\ref{fig:SpectrumFit} (d) as a function of DAQ date.

In the early days of the detector operation, its materials were already exposed to air, containing the radon leaking from the cave. It allows us to locate the $\beta$-$\alpha$ cascade decay of \ce{^{214}Bi}-\ce{^{214}Po}, which is the daughters of \ce{^{222}Rn}. After subtracting the accidental coincidences, the double coincidences in the first month of operation of the detector are the prompt-delayed signal pairs formed by $\beta$-$\alpha$. The distribution of its delayed signal energy is close to a Gaussian distribution, shown in Fig.~\ref{fig:Bi214EE} (c), which can also provide a supplementary information for the energy scale. In Fig.~\ref{fig:SpectrumFit} (d), the visible energy of \ce{^{214}Po}-$\alpha$ is taken as \SI{0.85}{\mega\electronvolt}, which is comparable to $\sim$\SI{0.9}{\mega\electronvolt} in Daya Bay experiment~\cite{DayaBay:2016ggj}.

It can be seen from Fig.~\ref{fig:SpectrumFit} (d) that the energy scale data of \ce{^{208}Tl}-$\gamma$ was always available during the detector operation, while \ce{^{40}K}-$\gamma$ and \ce{^{214}Po}-$\alpha$ were only available in the late and early stages of the detector operation, respectively. Therefore, we took the energy scale of \ce{^{208}Tl}-$\gamma$ as the central value for the time-dependent (approximately daily) energy scale for the whole period. While the energy scales of \ce{^{40}K}-$\gamma$ and \ce{^{214}Po}-$\alpha$ differed from those of \ce{^{208}Tl}-$\gamma$ by no more than $\sim$5\%, which was considered as systematic errors in the current analysis. Since the systematic error was much larger than the statistical error in the energy scale fittings, the latter was ignored in this study.

The convert factor used in energy scale calibration was then derived by taking the average in Fig.~\ref{fig:SpectrumFit} (d) for the different periods. This factor was determined to be \SI{61.07}{PE\per\mega\electronvolt} before July 14, 2019. It increased to \SI{88.02}{PE\per\mega\electronvolt} between July 15, 2019, and July 22, 2019. Finally, it was stable at \SI{99.85}{PE\per\mega\electronvolt} with an energy resolution of about $18\%/\sqrt{E}$\linelabel{line:energy resolution}. \linelabel{line:liquid-flow}This increasing trend indicates that oxygen is gradually removed with liquid convection during the degassing process. Compared to the current low PMT coverage of about $12\%$, the future detector will have a much higher light yield and better energy resolution with an increased coverage to about $80\%$.

\section{The detector simulation}
\label{sec:detector-simulation}

To understand the operation status of the 1-ton prototype, we built a simulation framework based on \texttt{GEANT4}~\cite{GEANT4:2002zbu}. In this paper, we mainly focused on the study of radioactive background level. The energy-related selection efficiencies and uncertainties for the backgrounds were therefore estimated with the simulated event samples.

The detector geometry and electronics parameters are customized via XML configurations. A streamed trigger mechanism is applied to emulate the actual events' coincidence and electronic trigger system. The natural radioactive decay chains of U, Th, K, Ac are fully considered. Their simulation generators are adapted from the NuDat database~\cite{NuDat} of the National Nuclear Data Center. The energy spectra of $\beta$ from the decays are calculated theoretically, including the corrections terms such as screening correction, finite-size correction, and weak magnetism correction. For short-lived nuclides, the time correlation of the cascade decay signals is also considered. It is particularly important for the \ce{^{214}Bi}-\ce{^{214}Po} and \ce{^{212}Bi}-\ce{^{212}Po} decay mentioned in Sec.~\ref{sec:measurement}.

The main components of the detector, including LS, acrylic vessel, PMTs, stainless steel tank, etc., are all implemented in the simulation according to the actual detector design. The optical simulation includes the \v{C}herenkov light and scintillation light. The gain of the PMT can vary during the different detector operating stages. We made real-time adjustments to it in the simulation based on the measurement results from the data. The TT and TTS were set to fixed values according to the test results of the PMTs.

\begin{figure}[!htb]
    \centering
    \subcaptionbox{Phase P1}{\includegraphics[width=0.49\textwidth]{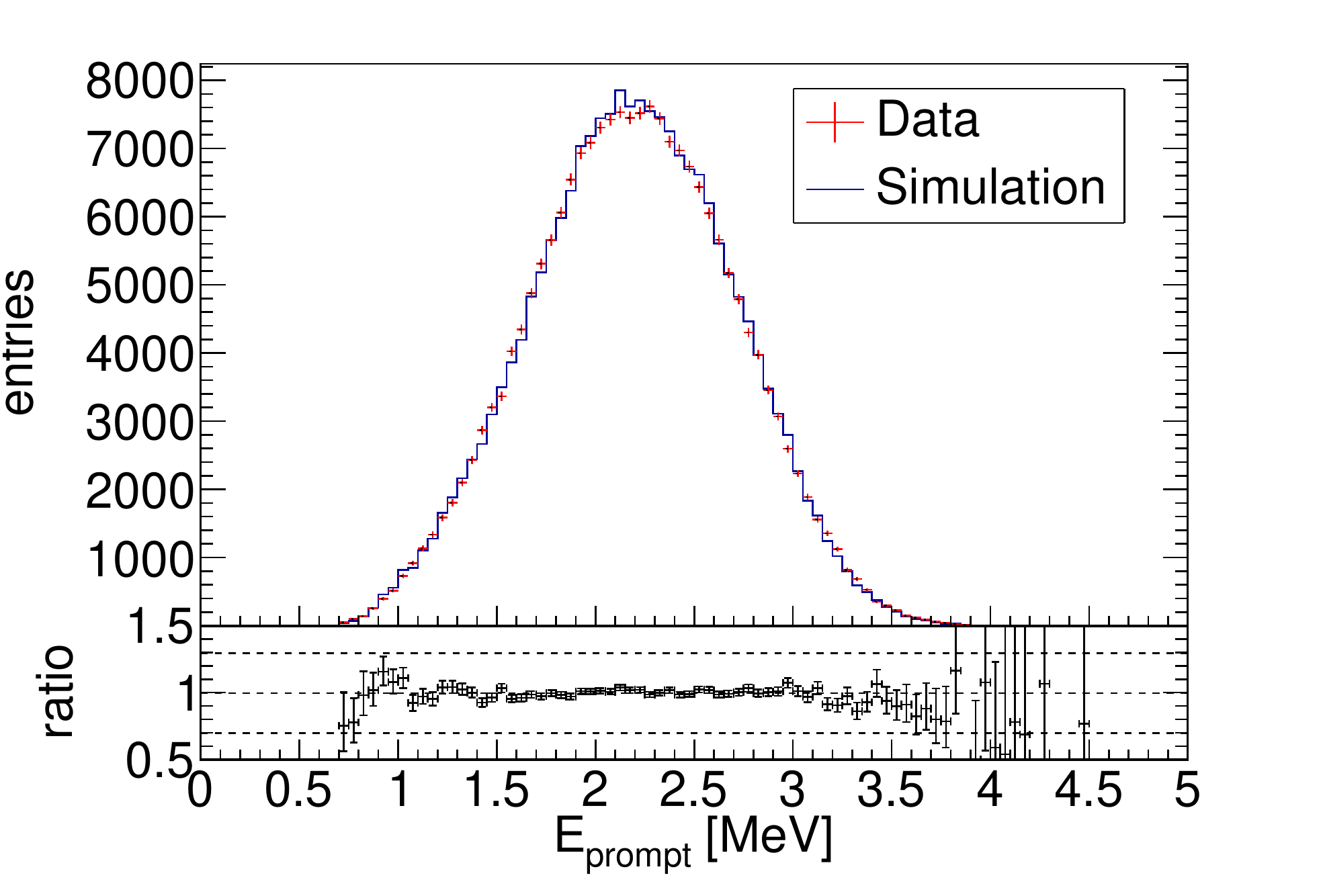}}
    \subcaptionbox{Phase P4}{\includegraphics[width=0.49\textwidth]{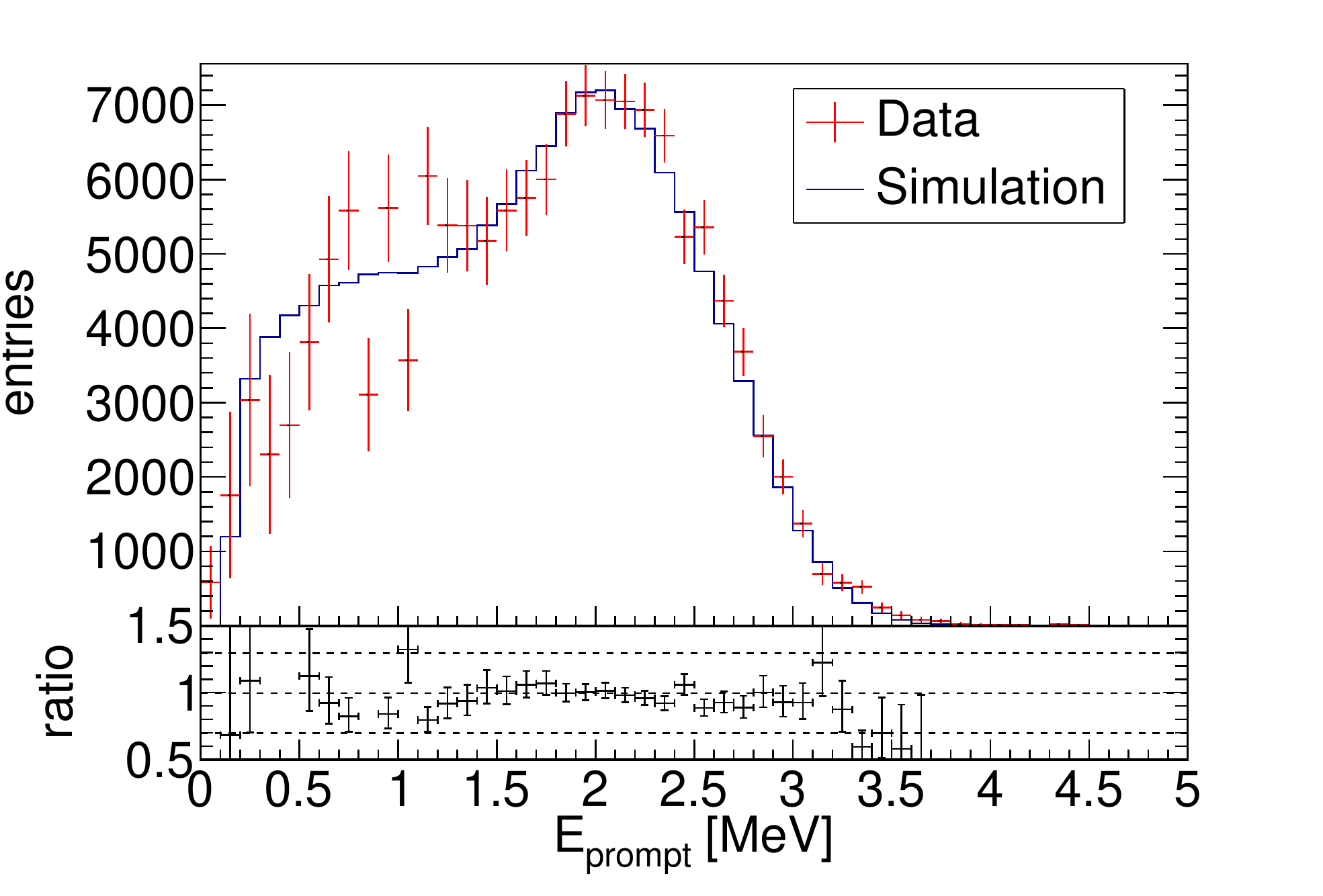}}
    \caption{Spectra comparison of \ce{^{214}Bi} decay between MC and data. The data spectra shown in Fig.~(a) and (b) are selected from P1 and P4 phases, respectively. The MC spectra are derived with the same detector running conditions as the data. The lower panel shows the bin-by-bin ratio of the two spectra.}
    \label{fig:Bi214BetaCompare}
\end{figure}

\linelabel{line:MC-validity}The validity of the detector simulation was confirmed with a comparison of the wide energy spectrum of \ce{^{214}Bi} decay between MC and data, as shown in Fig.~\ref{fig:Bi214BetaCompare}. We selected phase P1 and phase P4 to perform this comparison and found that the prompt spectrum of \ce{^{214}Bi}-\ce{^{214}Po} cascade decay in data agrees well with the MC one within the statistical error. The trigger conditions were significantly different for these two periods, including the threshold of the single channel, the required number of the fired PMTs, and the window length for recording a single triggered event. We implemented these changes in the detector simulation and got a consistent comparison result for both phases, which further verified the reliability of the detector simulation. This spectrum spans all interest energy ranges in this study, making this validity check rigorous enough. The difference in the low-energy tail of the spectrum between the two phases was due to the fact that we got much more low-energy triggers in P4 that in P1. To address the uncertainties of the selection efficiencies, we also compared the energy scale and resolution of $\alpha$ and $\gamma^\prime$s between MC and data. These differences were all taken into account when evaluating the efficiencies' uncertainty, which will be described in Sec.~\ref{sec:measurement}.

\section{Radioactive background analysis}
\label{sec:measurement}

We investigated the radioactive background from the LS and PMT using data collected with the 1-ton prototype.

\subsection{Event selection}
In the MeV scale, background signals currently found in the 1-ton prototype can be divided into two categories: uncorrelated single $\gamma$ and $\beta$-$\alpha$ signals from cascade decays. To study these signals from the data, we first performed a multiplicity selection by exploiting the time correlation between each event.

After the calibrations and data quality check, we selected events with an energy greater than \SI{0.2}{\mega\electronvolt}. The first such event initiates a coincidence time window of $T_c$=\SI{400}{\micro\second}. If a subsequent event occurs within the time window, they will be recorded as a coincidence event. If no subsequent event occurs within the time window, the event is recorded as an isolated event. Once the current window has ended, the first event falling outside it will initiate a new time window, and the process is repeated for all subsequent events. Finally, the events on the whole timeline were divided into isolated events and coincidence events. For the coincidence events, there are far more double coincidences than triple coincidences and others. For the double coincidence, the first and second events are called the prompt and delayed signal, respectively. Their reconstructed energies are denoted by $E_{\mathrm{p}}$ and $E_{\mathrm{d}}$, respectively.

\begin{figure}[!htb]
    \centering
    \includegraphics[width=0.7\linewidth]{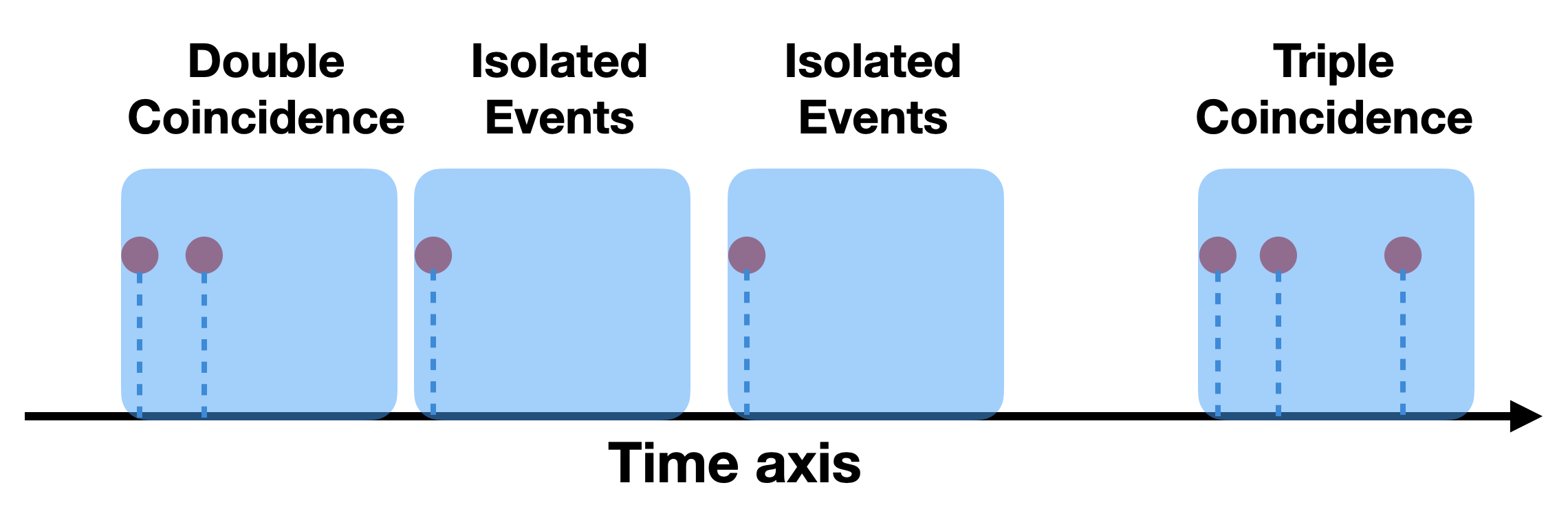}
    \caption{Schematic figure for the time coincidence of trigger events. The dark red dots represent trigger events. Blue rectangles are the coincidence time window with a fixed length $T_c$.}
    \label{fig:MultiplicitySelection}
\end{figure}

As an example illustrated in Fig.~\ref{fig:MultiplicitySelection}, there are a total of seven events satisfying $E>$\SI{0.2}{\mega\electronvolt} on the time axis, indicated by the red dots, and they are arranged in the order of their timestamps. The first event starts a time window, as shown by the light blue rectangle, and the time interval of the second event from it is less than $T_c$, so they are saved as a double coincidence. The time interval between the third event and the first event is greater than $T_c$, so the third event starts a new time window. Since there are no other events within the current time window, the third event is saved as an isolated event. The fourth event is similar to the third event and is also an isolated event. The fifth event starts a time window and is saved as a triple coincidence event along with the next two events.

After the multiplicity selection described above, we selected $\gamma$ signals from the isolated event sample and cascade decay signals from the double coincidence sample. The detailed selection criteria are shown in Tab.~\ref{tab:selection-cut}. The multiplicity is the number of events in the coincidence time window. For isolated events and double coincidences, the multiplicity is 1 and 2, respectively. $T_{\mathrm{pd}}$ is the time interval between the prompt and delayed signals. $D_{\mathrm{pd}}$ is the distance between the reconstructed vertices of the prompt and delayed signals.
\begin{table}[!htb]
    \caption{\label{tab:selection-cut}
        The selection criteria for the radioactive backgrounds in the 1-ton prototype. $T_{\mathrm{pd}}$ is the time interval between the prompt and delayed signals. $D_\mathrm{pd}$ is the distance between the prompt and delayed signal's reconstructed vertices.}
    \centering
    \begin{tabular}{ccccc}
        \midrule
        Quantity                                     & \ce{^{208}Tl}-$\gamma$ & \ce{^{40}K}-$\gamma$ & \ce{^{214}Bi}-\ce{^{214}Po} & \ce{^{212}Bi}-\ce{^{212}Po} \\
        \midrule
        Multiplicity                                 & 1                      & 1                    & 2                           & 2                           \\
        $T_{\mathrm{pd}}$ [\unit{\micro\second}]     & -                      & -                    & [1, 400]                    & [0.8, 10]                   \\
        $E_{\mathrm{p}}$ [\unit{\mega\electronvolt}] & [1.0, 2.0]             & [2.0, 3.5]           & [0.5, 3.5]                  & [0.2, 2.0]                  \\
        $E_{\mathrm{d}}$ [\unit{\mega\electronvolt}] & -                      & -                    & [0.4, 1.2]                  & [0.7, 1.0]                  \\
        $D_{\mathrm{pd}}$ [\unit{\milli\meter}]      & -                      & -                    & [0, 400]                    & [0, 400]                    \\
        \midrule
    \end{tabular}

\end{table}

In this study, the radioactive signal rates were determined with data collected in stage P4, which is after the deployment of the nitrogen bubbling system. The selected radioactive signal and the corresponding background rates are summarized in Tab.~\ref{tab:cut-eff}.

\subsubsection{\ce{^{214}Bi}-\ce{^{214}Po}}
\label{sec:BiPo214}
The half-life of \ce{^{214}Bi} is about \SI{19.9}{\minute}, and the branching ratio of releasing $\beta$ to decay to \ce{^{214}Po} is about 99.979\%~\cite{ENDF8}. The half-life of \ce{^{214}Po} is about \SI{164.3}{\micro\second}. Therefore, the emitted $\alpha$ by \ce{^{214}Po} has a strong time correlation with the $\beta$ from \ce{^{214}Bi} decay. After the above multiplicity selection, the \ce{^{214}Bi}-\ce{^{214}Po} cascade decay signal appears as a double coincidence composed of a prompt signal formed by \ce{^{214}Bi}-$\beta$ and a delayed signal formed by \ce{^{214}Po}-$\alpha$.

\begin{figure}[!htb]
    \centering
    \includegraphics[width=0.7\linewidth]{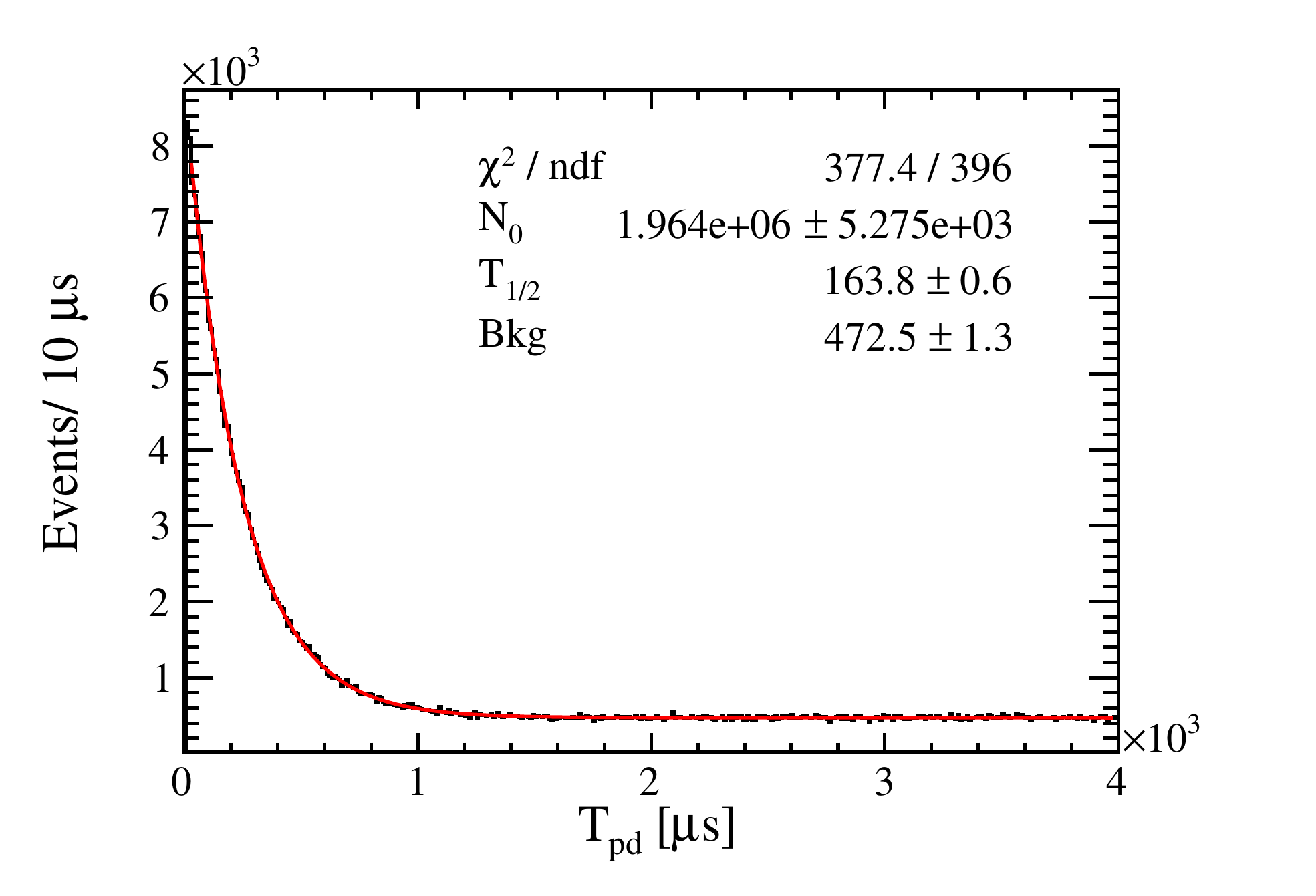}
    \caption{Distribution of time interval between prompt and delayed signals, after applying energy and distance cuts for \ce{^{214}Bi}-\ce{^{214}Po}. The red line shows the fitted function: $f(t)=\ln(2)N_0/T_{1/2}\exp\left(-\ln(2)t/T_{1/2}\right)+\mathrm{Bkg}$. $T_{1/2}$ represents the half-life of \ce{^{214}Po}. P1 data is used here for its large signal statistics.}
    \label{fig:Bi214TimeFit}
\end{figure}

Applying the cuts listed in Tab.~\ref{tab:selection-cut}, but adjusting the $T_{\mathrm{pd}}$ to \SI{4}{\milli\second}, we got a coincidence time distribution in a wide range using P1 data, as shown in Fig.~\ref{fig:Bi214TimeFit}. Fitting it with a function of exponential plus zero-order polynomial, we got a half-life of \SI{163.85(59)}{\micro\second}, which is in good agreement with \ce{^{214}Po}'s half-life listed in NuDat database~\cite{NuDat}, i.e., \SI{163.6}{\micro\second}.

The number of \ce{^{214}Bi}-\ce{^{214}Po} cascade decay signal was determined by two methods: 1. We fitted the time distribution between prompt and delayed signals using exponential function plus zero-order polynomial, then estimated the signal number by the exponential component; 2. We calculated the accidental background rate according to Ref.~\cite{Yu:2013cob}, then subtracted it from the \ce{^{214}Bi}-\ce{^{214}Po} candidates. The first method is straightforward.

For the second method, an accidental coincidence sample was constructed based on the selected isolated events during the same period. We took all the isolated events, a total of $N$, from a specified physical run. They were sorted by their timestamps. First, the $i$th in the first $N/2$ events was regarded as a prompt signal, and the $i$th in the last $N/2$ events was regarded as a delayed signal. They were combined as a pair-event. Then the $i$th of the last $N/2$ events was taken as the prompt signal, and the $i$th of the first $N/2$ events was taken as the delayed signal. They were combined again to improve the statistics of accidental sample. These event pairs need to satisfy the criteria listed in Tab.~\ref{tab:selection-cut}. To normalize the spectra of the accidental background, we first estimated the single event rate $R_s$ based on the number of isolated events and double coincidences. The accidental coincidence is a random combination of single events. Therefore, the accidental background rate can be precisely calculated with $R_s$ and $T_{\mathrm{c}}$ based on the principle described in Ref.~\cite{Yu:2013cob}.

\begin{figure}[!htb]
    \centering
    \includegraphics[width=0.7\linewidth]{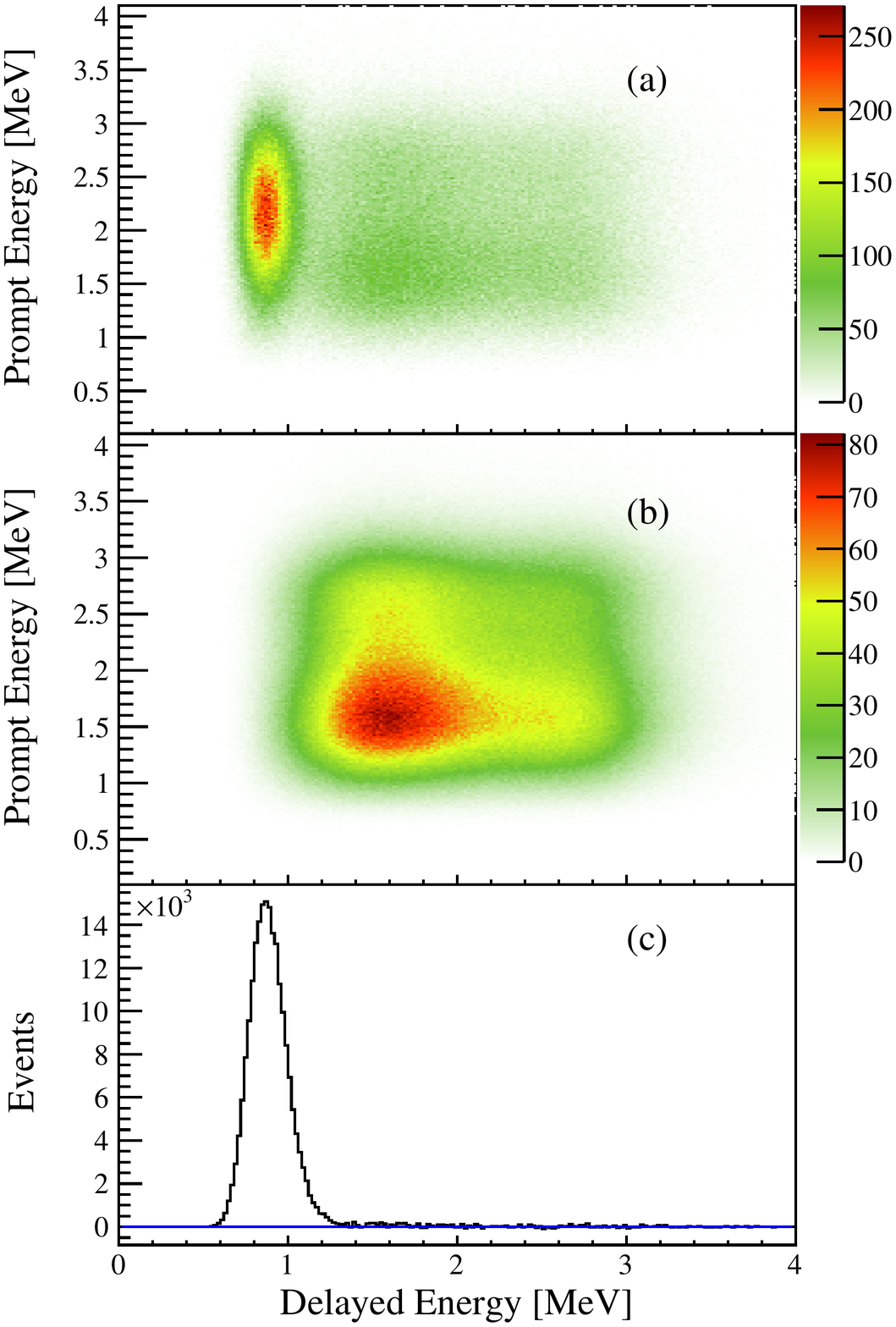}
    \caption{The prompt energy versus delayed energy for \ce{^{214}Bi}-\ce{^{214}Po} candidates (a) and accidental background (b). Fig.~(c) shows the delayed energy distribution after the background subtraction. P1 data is used here for its large signal statistics.}
    \label{fig:Bi214EE}
\end{figure}

Applying the nominal $T_{\mathrm{pd}}$ and $D_{\mathrm{pd}}$ cuts, but leaving $E_{\mathrm{p}}$ and $E_{\mathrm{d}}$ unlimited, Fig.~\ref{fig:Bi214EE} (a) demonstrates a clear correlation between the prompt and delayed signals of the \ce{^{214}Bi}-\ce{^{214}Po} cascade decay. Fig.~\ref{fig:Bi214EE} (b) shows the normalized energy distribution of the accidental background. The background-subtracted delayed energy spectrum is shown in Fig.~\ref{fig:Bi214EE} (c). The analyzed signal and background rates in P4 for \ce{^{214}Bi}-\ce{^{214}Po} candidates are listed in Tab.~\ref{tab:cut-eff}.

\begin{figure}[!htb]
    \centering
    \includegraphics[width=0.7\linewidth]{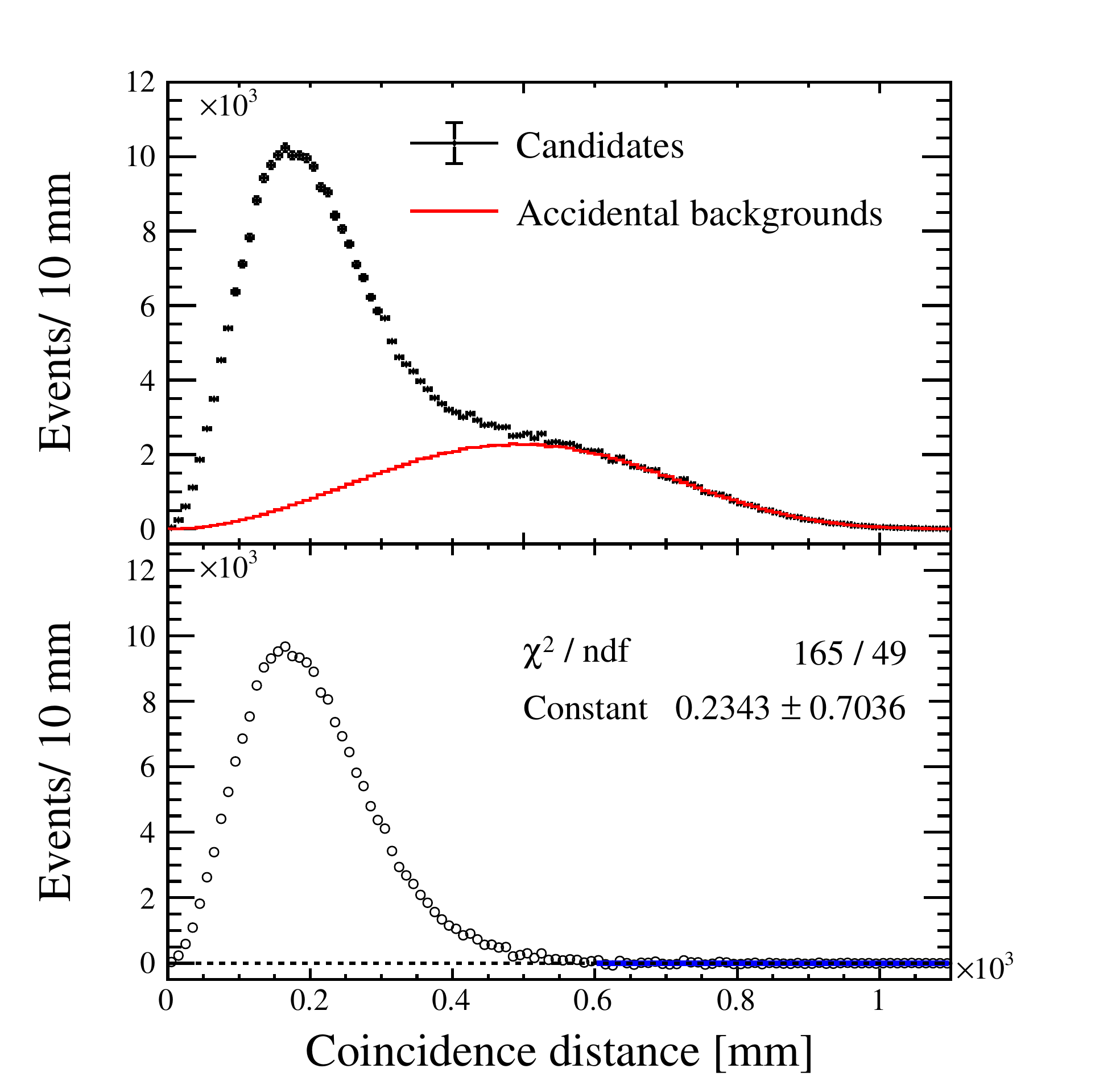}
    \caption{The upper panel shows the distance distributions between the reconstructed prompt and delayed vertices of the selected candidates and backgrounds. The lower panel shows the distance distribution of the candidates after background subtraction. The blue line shows the fit result of a zero-order polynomial in [600, 1100] mm.}
    \label{fig:Bi214D}
\end{figure}

To validate the background subtraction result, we adjusted the $D_{\mathrm{pd}}$ cut to 1.1 m. Fig.~\ref{fig:Bi214D} shows the distance distribution of \ce{^{214}Bi}-\ce{^{214}Po} candidates (black points) and accidental background (red line). The data collected in P1 was used here due to its large statistics. There are only accidental backgrounds when $D_{\mathrm{pd}}>$0.6 m. A fit of zero-order polynomial on the distance distribution with $D_{\mathrm{pd}}>$0.6 m shows a good agreement with Zero statistically, as shown in the lower panel of Fig.~\ref{fig:Bi214D}. This indicates the validity of the background subtraction.

To investigate the source of \ce{^{214}Bi}-\ce{^{214}Po} signal, we analyzed the evolution in its signal number over date. The number of \ce{^{214}Bi}-\ce{^{214}Po} signals showed an exponential decrease in stage P1, indicating that the LS was mixed with radon during the process of filling the detector in the experimental hall. \ce{^{214}Bi} is the daughter of \ce{^{222}Rn}, and its half-life is much smaller than that of \ce{^{222}Rn}. Fitting the trend in \ce{^{214}Bi}-\ce{^{214}Po} signal rate yielded a half-life of \SI{3.95(07)}{\days}, which is consistent with that of \ce{^{222}Rn}.
\begin{figure*}[!htb]
    \centering
    \includegraphics[width=0.7\linewidth]{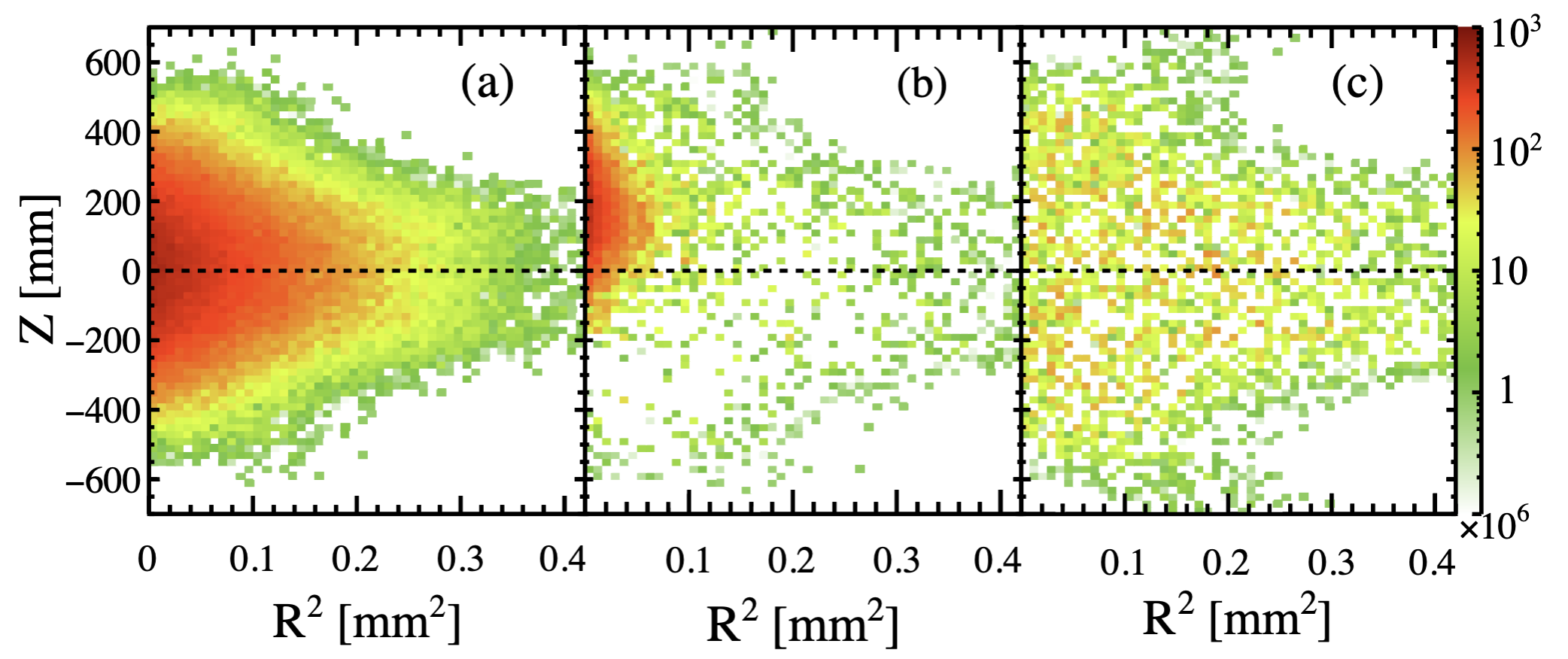}
    \caption{The reconstructed vertex distribution of \ce{^{214}Bi}-\ce{^{214}Po} signal after the background subtraction. Fig.~(a), (b) and (c) correspond to the DAQ phase P1, P2+P3 and P4 respectively.}
    \label{fig:Bi214vertex}
\end{figure*}

Figure \ref{fig:Bi214vertex} further shows the vertex distribution of \ce{^{214}Bi}-\ce{^{214}Po} signal in different operating stages. Below is an explanation of the characteristics of the three graphs in Fig.~\ref{fig:Bi214vertex} at different stages.
\begin{itemize}
    \item \textbf{Dissolved Radon Before Detector Installation}. In the P1 phase, the radon gas was already dissolved into the LS before being filled into the detector. Therefore, the vertex distribution in Fig.~\ref{fig:Bi214vertex} (a) is up-down symmetrical in the detector.
    \item \textbf{Radon Leakage from Chimney}. In the P2 and P3 phases, the original dissolved radon gas had decayed, so the \ce{^{214}Bi}-\ce{^{214}Po} signal was dominated by the leaked radon from the chimney at the top of the detector. So the vertex distribution in Fig.~\ref{fig:Bi214vertex} (b) shows a clustering at the top of the detector.
    \item \textbf{Intrinsic \ce{^{238}U} decay after Nitrogen Sealing}. In the P4 phase, to suppress the radon leakage~\cite{Zuzel:2018ngj}, we installed a nitrogen bubbling system. The \ce{^{214}Bi}-\ce{^{214}Po} signal in P4 stage mainly came from the intrinsic natural \ce{^{238}U} decay chain in LS itself. Therefore, the vertex distribution in Fig.~\ref{fig:Bi214vertex} (c) is also up-down symmetrical.
\end{itemize}

The installation of the nitrogen bubbling system suppressed the radon leakage. It reduced the \ce{^{214}Bi}-\ce{^{214}Po} signal rate by about 4.8 times. Note that the vertex distribution in Fig.~\ref{fig:Bi214vertex} is background-subtracted, so some bins may have negative content due to statistical fluctuations. To bring the three pictures to a unified $z$ range, we chose the log scale, so the bins less than 0 are displayed as white in Fig.~\ref{fig:Bi214vertex}. This is particularly severe for Fig.~\ref{fig:Bi214vertex} (c). Because of the small number of signals in the P4 phase, there are significant statistical fluctuations almost everywhere.

\subsubsection{\ce{^{208}Tl}-$\gamma$}
\ce{^{208}Tl} can decay to the first excited state of \ce{^{208}Pb} by emitting a $\beta$. During its deexcitation, a $\gamma$ of \SI{2.61}{\MeV} will be emitted, and the timescale is picoseconds. For \ce{^{208}Tl} in the LS, the monoenergetic $\gamma$, and $\beta$ of continuous energy are mixed and cannot be distinguished. In the data, it corresponds to an energy distribution of around \SI{3.8}{\MeV}, which hasn't been observed. For \ce{^{208}Tl} in the PMTs, steel frame, and other structures of the prototype, the emitted $\beta$ is shielded by the water layer and cannot enter into LS. Therefore, \ce{^{208}Tl}-$\gamma$ emitted from the PMTs can be studied.

In the data, the background source in the isolated event sample is complex. We can't subtract them like the \ce{^{214}Bi} analysis. We simply describe the \ce{^{208}Tl}-$\gamma$ peak with a Gaussian distribution. The backgrounds are approximated with an exponential distribution. Since there is virtually no distribution higher than this peak, the right side of the Gaussian distribution can be well constrained. The signal-to-noise ratio at the peak is relatively good, about 2. Therefore, it is reliable to fit the energy peak with Eq.~\ref{eq:fit-Tl208}. In this study, the number of \ce{^{208}Tl}-$\gamma$ signal was determined a fit using Eq.~\ref{eq:fit-Tl208}.
\begin{equation}
    \label{eq:fit-Tl208}
    f_{\mathrm{Tl}}(E)=\frac{N_{\mathrm{Tl}}}{\sqrt{2 \pi} \sigma_{\mathrm{Tl}}} \exp \left(-\frac{(E-\mu_{\mathrm{Tl}})^2}{2 \sigma_{\mathrm{Tl}}^2}\right)+f_\mathrm{bkg},
\end{equation}
where $N_{\mathrm{Tl}}$ is the signal number of \ce{^{208}Tl},
$\mu_{\mathrm{Tl}}$ and $\sigma_{\mathrm{Tl}}$ are the peak position and resolution of \ce{^{208}Tl}-$\gamma$, $f_\mathrm{bkg}$ is the energy distribution approximated with the exponential functions.

Figure \ref{fig:SpectrumFit} (a)-(c) show the typical energy spectrum of the isolated events, where the pink line shows the best fit result of the Gaussian component in Eq.~\ref{eq:fit-Tl208}. The analyzed signal and background rates in P4 for \ce{^{208}Tl}-$\gamma$ candidates are listed in Tab.~\ref{tab:cut-eff}.

\subsubsection{\ce{^{40}K}-$\gamma$}
\ce{^{40}K} can decay to the first excited state of \ce{^{40}Ar} through orbital electron capture with a branch ratio of $10.86\%$~\cite{ENDF8}. The \SI{1.46}{\MeV} $\gamma$ emitted during its deexcitation can be used to label \ce{^{40}K} signal.

The energy peak of \ce{^{40}K}-$\gamma$ is lower than that of \ce{^{208}Tl}-$\gamma$, suffering from more radioactive backgrounds. When fitting its energy spectrum, we first fixed the \ce{^{208}Tl}-$\gamma$ peak, as shown in Eq.~\ref{eq:fit-Tl208}. The fit function is given in Eq.~\ref{eq:fit-K40}.
\begin{multline}\label{eq:fit-K40}
    f_{\mathrm{K}}(E)=\frac{N_{\mathrm{K}}}{\sqrt{2 \pi} \sigma_{\mathrm{K}}} \exp \left(-\frac{(E-\mu_{\mathrm{K}})^2}{2 \sigma_{\mathrm{K}}^2}\right)+\frac{N_{\mathrm{Tl}}}{\sqrt{2 \pi} \sigma_{\mathrm{Tl}}} \exp \left(-\frac{(E-\mu_{\mathrm{Tl}})^2}{2 \sigma_{\mathrm{Tl}}^2}\right)+f_\mathrm{bkg},
\end{multline}
where $N_{\mathrm{K}}$ is the signal number of \ce{^{40}K},
$\mu_{\mathrm{K}}$ and $\sigma_{\mathrm{K}}$ are the peak position and resolution of \ce{^{40}K}-$\gamma$, $f_\mathrm{bkg}$ is the energy distribution approximated with the exponential functions. Other parameters in Eq.~\ref{eq:fit-K40} are fixed at the best-fit values of \ce{^{208}Tl}-$\gamma$. The typical fitting result is shown in Fig.~\ref{fig:SpectrumFit} (b) and (c), where the blue lines show the fitted \ce{^{40}K}-$\gamma$ peaks. The analyzed signal and background rates in P4 for \ce{^{40}K}-$\gamma$ candidates are listed in Tab.~\ref{tab:cut-eff}.

\subsubsection{\ce{^{212}Bi}-\ce{^{212}Po}}\label{sec:bi212-analysis}
The signature of \ce{^{212}Bi}-\ce{^{212}Po} cascade decay is basically similar to that of \ce{^{214}Bi}-\ce{^{214}Po}, except that \ce{^{212}Po} has a shorter half-life of about \SI{300}{\nano\second}. Consequently, it gives a much smaller time interval between the prompt and delayed signals.

The procedure to get the number of \ce{^{212}Bi}-\ce{^{212}Po} events should be similar to the \ce{^{214}Bi} analysis. But our study showed the contamination level of \ce{^{232}Th} can be much lower than that of \ce{^{238}U}. The scintillation light of slow LS lasts longer. It may also complicate the detection of highly short-lived cascade decays compared to the LS. The photons produced by \ce{^{212}Bi}-$\beta$ in LS can be aliased in time with those produced by \ce{^{212}Po}-$\alpha$.

\begin{figure}[!htb]
    \centering
    \includegraphics[width=0.7\linewidth]{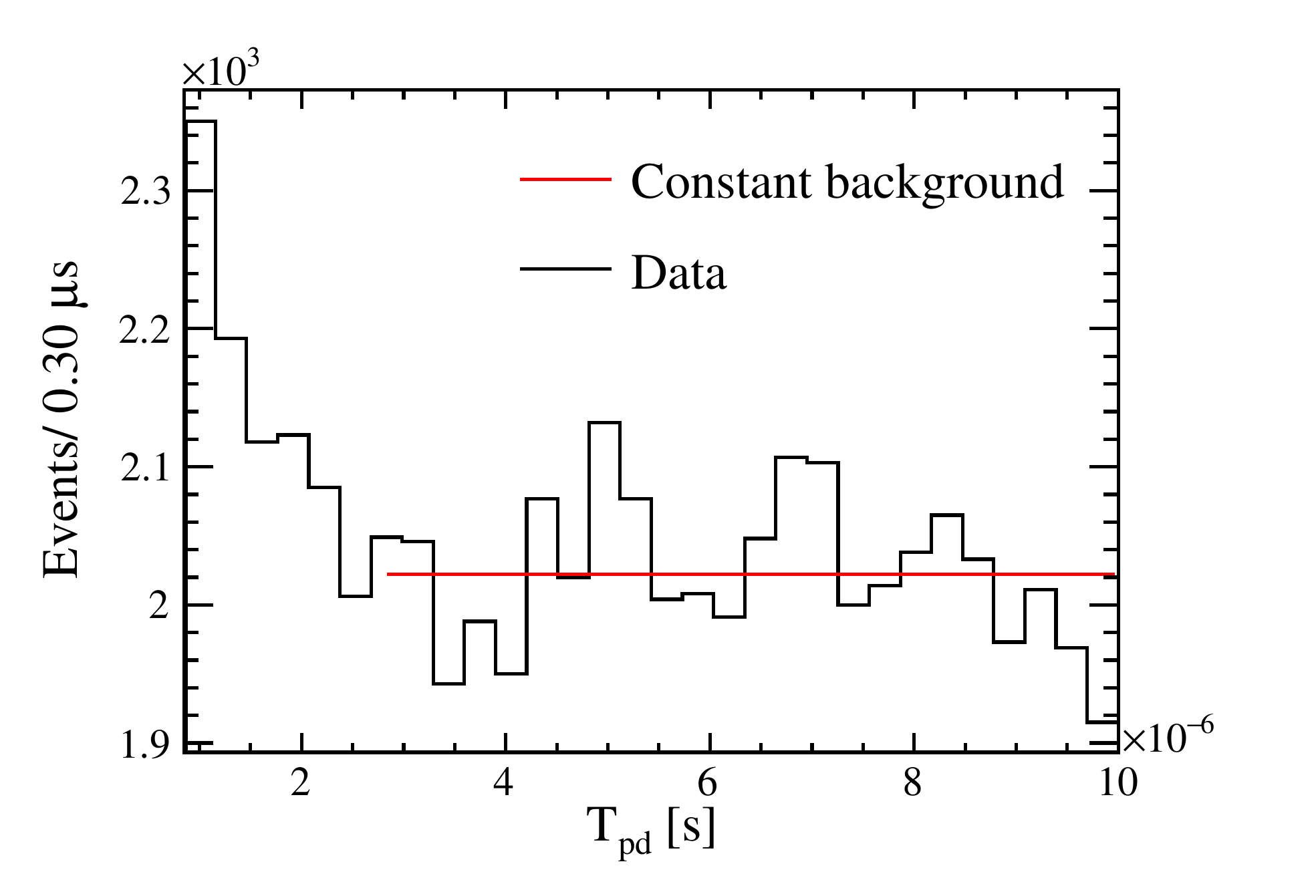}
    \caption{Distribution of time interval between prompt and delayed signals, after applying energy and distance cuts for \ce{^{212}Bi}-\ce{^{212}Po}. The red line shows the fit result of a constant function to the data beyond \SI{2.8}{\micro\second}, which provides the estimation of accidental background in \ce{^{212}Bi}-\ce{^{212}Po} candidates.}
    \label{fig:Bi212Dt}
\end{figure}

\linelabel{line:bi212-fig-begin}
After applying the \ce{^{212}Bi}-\ce{^{212}Po} selection criteria shown in Tab.~\ref{tab:selection-cut}, we obtained the time distribution between prompt and delayed signals as shown in Fig.~\ref{fig:Bi212Dt}. Significantly more events below \SI{2.8}{\micro\second} were the \ce{^{212}Bi}-\ce{^{212}Po} candidates. However, these events also contained some after-pulses, and current analysis methods cannot distinguish them due to the limit of statistics. The energy distributions in that time interval were also checked to make sure here is no visible signal peaks. The time distribution of the candidates beyond \SI{2.8}{\micro\second} was uniform, and it is known to be the contribution of accidental coincidences. We determined the number of accidentals by fitting the uniform distribution, as shown in the red line in Fig.~\ref{fig:Bi212Dt}, and extrapolating the result to the sample below \SI{2.8}{\micro\second}. Therefore, in this study, we assumed that the events with $T_{\mathrm{pd}}<$ \SI{2.8}{\micro\second} in Fig.~\ref{fig:Bi212Dt} are all \ce{^{212}Bi}-\ce{^{212}Po} signals, and conservatively estimated the maximal decay rate of \ce{^{212}Bi} in the detector.
\linelabel{line:bi212-fig-end} The analyzed background rate and upper limit of signal rate in P4 for \ce{^{212}Bi}-\ce{^{212}Po} candidates are listed in Tab.~\ref{tab:cut-eff}.

\subsection{Selection efficiencies and uncertainties}
The decay rates of \ce{^{214}Bi}, \ce{^{208}Tl}, \ce{^{40}K} and \ce{^{212}Bi} in the 1-ton prototype can be determined by Eq.~\ref{eq:background-number}.
\begin{equation}\label{eq:background-number}
    Y=\dfrac{R_{\mathrm{sig}}}{\varepsilon_m\cdot\varepsilon_\mathrm{ep}\cdot\varepsilon_\mathrm{ed}\cdot\varepsilon_\mathrm{tc}\cdot\varepsilon_\mathrm{dc}\cdot M \cdot B}
\end{equation}
where $R_{\mathrm{sig}}$ is the selected signal rate, $\varepsilon_m$ is the multiplicity selection efficiency, $\varepsilon_\mathrm{ep},~\varepsilon_\mathrm{ed},~\varepsilon_\mathrm{tc}$, and $\varepsilon_\mathrm{dc}$ are the selection efficiency for $E_{\mathrm{p}}$, $E_{\mathrm{d}}$, $T_{\mathrm{pd}}$ and $D_{\mathrm{pd}}$ respectively, $M$ is the mass of the material containing the corresponding nuclide, $B$ is the branching ratio of this nuclide to form the corresponding signal.

The multiplicity selection efficiency can be calculated precisely based on the data. The two single events may be randomly combined to form an accidental coincidence, so that the detected single signal number, i.e., the isolated event number, is smaller than the actual number of single signals before the event selection. For a double coincidence, the correlated prompt and delayed signals may be interrupted by a single signal, which may be a single signal appearing before the prompt signal within $T_c$, or between the prompt and delayed signals, or after the delayed signal to form a triple coincidence. In this study, $\varepsilon_m$ was evaluated to be about 0.8818. The value of $\varepsilon_m$ depends on the estimation of the single event rate (about \SI{160}{\hertz}), but the relative variation in the latter of no more than 5\% over time, which can only result in about a 0.6\% difference in $\varepsilon_m$ over time. For each physics run, the statistical error of the $\varepsilon_m$ was no more than $0.3\%$ due to the large statistics of the isolated event sample. Therefore, its uncertainty is negligible in this study.

The mass of LS was determined to be \SI{955.4}{\kilogram} with negligible error. The mass of PMT glass was estimated to be \SI{770}{\gram} with a 10\% relative error, based on measured PMT data with the same model. In this paper, we assumed that \ce{^{208}Tl}-$\gamma$ and \ce{^{40}K}-$\gamma$ both come from the PMTs, \ce{^{212}Bi}-\ce{^{212}Po} and \ce{^{214}Bi}-\ce{^{214}Po} both came from the LS.

The value of $B$ is taken from the public database~\cite{ENDF8}. For \ce{^{214}Bi}-\ce{^{214}Po} and \ce{^{212}Bi}-\ce{^{212}Po}, $B$ is $99.979\%$ and $64.06\%$, respectively, which are the products of $\beta$ and $\alpha$ decays branch ratios. For \ce{^{208}Tl}-$\gamma$ and \ce{^{40}K}-$\gamma$, $B$ is $100\%$ and $10.86\%$, respectively, representing their probability of forming \SI{2.61}{\MeV} $\gamma$ and \SI{1.46}{\MeV} $\gamma$ signals respectively. Note that if the selected \ce{^{208}Tl} is the decay product of \ce{^{232}Th}, there is a branch ratio of about $35.94\%$ from \ce{^{232}Th} to \ce{^{208}Tl}.

For \ce{^{208}Tl}-$\gamma$ and \ce{^{40}K}-$\gamma$, only $\varepsilon_m$ and $\varepsilon_{\mathrm{ep}}$ listed in Eq.~\ref{eq:background-number} are effective. Their nominal values of $\varepsilon_{\mathrm{ep}}$ were estimated with MC samples. For \ce{^{212}Bi}-\ce{^{212}Po} and \ce{^{214}Bi}-\ce{^{214}Po} signals, $\varepsilon_{\mathrm{ep}}$ and $\varepsilon_{\mathrm{ed}}$ were also estimated with MC samples. To evaluate $\varepsilon_{\mathrm{tc}}$ and $\varepsilon_{\mathrm{dc}}$, the complete distance and time distribution were derived for \ce{^{214}Bi}-\ce{^{214}Po} based on data by adjusting the distance and time cuts to \SI{2}{\meter} and \SI{1500}{\micro\second}, respectively. The nominal value of $\varepsilon_{\mathrm{dc}}$ ($\varepsilon_{\mathrm{tc}}$) was therefore directly calculated by the ratio of integral of distance (time) distribution in the cut region and whole region. For the time distribution of \ce{^{214}Bi}-\ce{^{214}Po}, the efficiency loss caused by \SI{1}{\micro\second} and \SI{1500}{\micro\second} cut was evaluated to be about 0.6\% theoretically. The $\varepsilon_{\mathrm{tc}}$ used in this study was already corrected for this effect. Due to the absence of \ce{^{212}Bi}-\ce{^{212}Po} in the data, The nominal values of $\varepsilon_{\mathrm{dc}}$ and $\varepsilon_{\mathrm{tc}}$ are both estimated with MC samples. These efficiencies are summarized together with signal rate $R_{\mathrm{sig}}$ and background rate $R_{\mathrm{bkg}}$ in Tab.~\ref{tab:cut-eff}.
\begin{table}[!htb]
    \caption{\label{tab:cut-eff}
        The estimated selection efficiencies for the radioactive signals, and selected signal rate $R_{\mathrm{sig}}$ and background rate $R_{\mathrm{bkg}}$.}
    \centering
    \begin{tabular}[b]{ccccc}
        \midrule
        Quantity                  & \ce{^{208}Tl}-$\gamma$ & \ce{^{40}K}-$\gamma$ & \ce{^{214}Bi}-\ce{^{214}Po} & \ce{^{212}Bi}-\ce{^{212}Po} \\
        \midrule
        $R_{\mathrm{sig}}$        & \SI{3.60}{\hertz}      & \SI{3.04}{\hertz}    & 783.96/day                  & $<$4.74/day                 \\
        $R_{\mathrm{bkg}}$        & \SI{1.80}{\hertz}      & \SI{47.96}{\hertz}   & 36413.96/day                & 75.37/day                   \\
        $\varepsilon_\mathrm{tc}$ & -                      & -                    & 0.83                        & 0.15                        \\
        $\varepsilon_\mathrm{ep}$ & 0.11                   & 0.11                 & 0.87                        & 0.91                        \\
        $\varepsilon_\mathrm{ed}$ & -                      & -                    & 0.97                        & 0.72                        \\
        $\varepsilon_\mathrm{dc}$ & -                      & -                    & 0.97                        & 0.98                        \\
        \midrule
    \end{tabular}
\end{table}

The $R_{\mathrm{sig}}$ for \ce{^{208}Tl}-$\gamma$ and \ce{^{40}K}-$\gamma$ were determined by the Gaussian integrals within the respective energy ranges listed in Tab.~\ref{tab:selection-cut}. The fit errors induced uncertainties are negligible in this work. For \ce{^{214}Bi}-\ce{^{214}Po}, $R_{\mathrm{sig}}$ was determined by two independent methods, of which the first method gave the central value with a relative uncertainty of about $2\%$. The difference in the results between the two methods was about $10.2\%$. The combined uncertainty of $R_{\mathrm{sig}}$ is therefore $10.4\%$ in this study. The uncertainties of $R_{\mathrm{sig}}$ for \ce{^{212}Bi}-\ce{^{212}Po} was dominated by the statistic error of the accidental background, which was determined by the fitted constant in Fig.~\ref{fig:Bi212Dt}.

For \ce{^{208}Tl}-$\gamma$ and \ce{^{40}K}-$\gamma$, the fitted mean and sigma of the Gaussian peak was the source of the $\varepsilon_{\mathrm{ep}}$ efficiency uncertainties. The energy scale uncertainty was evaluated to be about 5\% in the previous calibration section. We varied the energy scale of MC events by the same size of uncertainty to investigate the variation of $\varepsilon_{\mathrm{ep}}$. While the uncertainty of fitted Gaussian sigma was determined by the comparison of that in data and MC. The fitted Gaussian sigma varied with the DAQ date, of which the maximal difference from that in MC is about 27\%. Varying the Gaussian sigma in the MC sample by the same amount of uncertainty can give the estimation of the efficiency uncertainties induced by resolution. The final total uncertainty of $\varepsilon_{\mathrm{ep}}$ is evaluated to be about 27\% for \ce{^{208}Tl}-$\gamma$ in this study. The procedure for estimating $\varepsilon_{\mathrm{ep}}$ for \ce{^{40}K}-$\gamma$ is similar. The relative uncertainties of $\varepsilon_{\mathrm{ep}}$ for \ce{^{208}Tl}-$\gamma$ and \ce{^{40}K}-$\gamma$ are summarized in Tab.~\ref{tab:uncertainties}.

The uncertainties of $\varepsilon_{\mathrm{ep}}$ and $\varepsilon_{\mathrm{ed}}$ for \ce{^{214}Bi}-\ce{^{214}Po} were dominated by the energy scale uncertainty, estimated with MC samples. The uncertainty of the time cut efficiency was given by the difference between the data and theoretical calculation. In the latter, we directly took the time distribution of \ce{^{214}Bi}-\ce{^{214}Po} as a standard exponential distribution. Using of the large statistics of \ce{^{214}Bi}-\ce{^{214}Po} in the first month of the detector running, we estimated the uncertainty of the distance cut efficiency to be about $0.2\%$, which is negligible in this work.

The uncertainties of $\varepsilon_{\mathrm{ep}}$ and $\varepsilon_{\mathrm{ed}}$ for \ce{^{212}Bi}-\ce{^{212}Po} were evaluated in the same way as \ce{^{214}Bi}-\ce{^{214}Po}. However, the uncertainty of the time and distance cut efficiencies can only be studied with the MC sample. The uncertainty of $\varepsilon_{\mathrm{tc}}$ was dominated by the MC statistics and the difference between the value derived from the MC sample and theoretical calculation. The uncertainty of $\varepsilon_{\mathrm{dc}}$ was dominated by the MC statistics, and the difference between the data and MC. The latter was taken from the \ce{^{214}Bi}-\ce{^{214}Po} study. The combined uncertainty is still less than $1\%$ in this work, which is negligible.

\begin{table}[!htb]
    \centering
    \caption{\label{tab:uncertainties}
        The estimated relative uncertainties for $R_{\mathrm{sig}}$, $\varepsilon_\mathrm{tc}$, $\varepsilon_\mathrm{ep}$, and $\varepsilon_\mathrm{ed}$ in this study for different radioactive signals. The evaluated uncertainties of $\varepsilon_{\mathrm{dc}}$ are less than 1\% for both \ce{^{214}Bi}-\ce{^{214}Po} and \ce{^{212}Bi}-\ce{^{212}Po}, thus ignored in this study.}
    \begin{tabular}{ccccc}
        \midrule
                                  & \ce{^{208}Tl}-$\gamma$ & \ce{^{40}K}-$\gamma$ & \ce{^{214}Bi}-\ce{^{214}Po} & \ce{^{212}Bi}-\ce{^{212}Po} \\
        \midrule
        $R_{\mathrm{sig}}$        & -                      & -                    & 10.4\%                      & 16.0\%                      \\
        $\varepsilon_\mathrm{tc}$ & -                      & -                    & 1.2\%                       & 7.9\%                       \\
        $\varepsilon_\mathrm{ep}$ & 26.9\%                 & 27.0\%               & 7.3\%                       & 1.9\%                       \\
        $\varepsilon_\mathrm{ed}$ & -                      & -                    & 1.5\%                       & 8.8\%                       \\
        \midrule
    \end{tabular}
\end{table}

All uncertainties that contribute to the final measurement results are summarized in Tab.~\ref{tab:uncertainties}. The systematic uncertainties of less than $1\%$ are ignored in this study.

\subsection{Analysis results}

Based on the event selections and efficiency studies, we measured the decay rate of \ce{^{208}Tl}, \ce{^{40}K} and \ce{^{214}Bi}, and estimated the upper limit of \ce{^{212}Bi} in the 1-ton prototype. Assuming long-term equilibrium holds, we gave the measurements or upper limits of the contamination levels for the \ce{^{238}U}, \ce{^{232}Th}, and \ce{^{40}K} decay chains, as listed in Tab.~\ref{tab:radio-measurement}. In the lower part of the table, the amounts of \ce{^{232}Th} in the PMT and LS were converted from the quantities of \ce{^{208}Tl} and \ce{^{212}Bi} listed in the upper part of the table, while the value of \ce{^{238}U} in LS was from the quantities of \ce{^{214}Bi}.

\begin{table*}[tp]
    \centering
    \caption{
        Decay rates of \ce{^{208}Tl}, \ce{^{40}K}, \ce{^{214}Bi}, and upper limit of \ce{^{212}Bi}'s decay rate in the PMT and LS. Under the long-term equilibrium assumption, the converted contamination levels are also shown in the lower part of this table.
    }
    \label{tab:radio-measurement}
    \sisetup{exponent-mode=scientific}
    \begin{tabular}{cccc}
        \midrule
                                                 &               & PMT                 & LS                     \\
        \midrule
        \multirow{4}*{Decay rate [Bq/g]}         & \ce{^{214}Bi} & -                   & \num{1.59(0.20)e-8}    \\
                                                 & \ce{^{208}Tl} & \num{1.64(0.47)e-3} & -                      \\
                                                 & \ce{^{212}Bi} & -                   & $<$\num{1.01(0.20)e-9} \\
                                                 & \ce{^{40}K}   & \num{1.24(35)e-2}   & -                      \\
        \midrule
        \multirow{3}*{Contamination level [g/g]} & \ce{^{238}U}  & -                   & \num{1.28(16)e-12}     \\
                                                 & \ce{^{232}Th} & \num{1.12(32)e-6}   & $<$\num{2.49(50)e-13}  \\
                                                 & \ce{^{40}K}   & \num{4.67(1.35)e-8} & -                      \\
        \midrule
    \end{tabular}
\end{table*} 
\section{Summary and discussion}
\label{sec:summary}

This paper reported the performance of the 1-ton prototype of the Jinping Neutrino
Experiment (JNE). The completed data analysis pipeline has been established, including the gain and time calibration of the PMTs, the vertex reconstruction, and the simulation framework. We observed different characteristics of \ce{^{222}Rn} background from various sources in the prototype, and used nitrogen sealing technology to control it, reducing the internal \ce{^{214}Bi}-\ce{^{214}Po} decay rate by about 4.8 times. We finally measured the contamination levels of \ce{^{238}U}, \ce{^{232}Th}, and \ce{^{40}K} in the detector materials.

\linelabel{line:1t-performance}This prototype verified the application of the hardware equipment and technology and gave plenty of information about the backgrounds. \linelabel{line:compare radioactivity}The contamination levels of \ce{^{238}U} and \ce{^{232}Th} are three more orders lower than the $10^{-16}$ g/g or less required by solar neutrino physics study in the MeV energy scale. \linelabel{line:promise-tech}Further studies on background suppression are still underway by adopting the techniques used in KamLAND~\cite{KamLAND2009PhDT} and Borexino~\cite{Zuzel:2015wla,BOREXINO:2018ohr}\linelabel{line:promise-tech-end}. \linelabel{line:1t upgrade plan}In addition, due to the limit of coverage, the feature of slow LS has not been exploited yet, for example, the fast and slow components of light for charged particles in large-scale detectors. \linelabel{line:1t-pmt-doubling}We expect an ongoing upgrade of this prototype can demonstrate the ability of slow LS by doubling the coverage of PMTs and using those with higher detection efficiency, lower background glass, and better time feature.\linelabel{line:1t-pmt-doubling-end} The study will advance the JNE with multi-hundred-ton target material, which is now under construction at CJPL-II.

\section{Acknowledgements}
This work was supported in part by the National Natural Science Foundation of China (12127808, 12141503) and the Key Laboratory of Particle and Radiation Imaging (Tsinghua University). We acknowledge Orrin Science Technology, Jingyifan Co., Ltd, and Donchamp Acrylic Co., Ltd, for their efforts in the engineering design and fabrication of the stainless steel and acrylic vessels. Many thanks to the CJPL administration and the Yalong River Hydropower Development Co., Ltd. for logistics and support.

\bibliographystyle{unsrt}
\bibliography{bibfile}

\end{document}